\documentclass[useAMS,usenatbib]{mn2e}

\usepackage{latexsym} 

\title[Evolution of QSO host galaxies]
{The evolution of star formation in quasar host galaxies}
\author[Stephen Serjeant and Evanthia Hatziminaoglou]
{Stephen Serjeant$^{1}$ and Evanthia Hatziminaoglou$^{2}$\\
$^{1}$Department of Physics \& Astronomy, Venables Building, The Open
University, Milton Keynes, MK7 6AA, UK\\
$^{2}$European Southern Observatory, Karl-Schwarzschild-Str. 2, 85748
Garching bei M\"{u}nchen, Germany
}
\begin{document}

\input BoxedEPS.tex
\SetEPSFDirectory{./}

\SetRokickiEPSFSpecial
\HideDisplacementBoxes

\date{Accepted; Received; in original form}

\pagerange{\pageref{firstpage}--\pageref{lastpage}} \pubyear{2008}

\maketitle

\label{firstpage}

\begin{abstract}
  We have used far-infrared data from IRAS, ISO, SWIRE, SCUBA and
  MAMBO to constrain statistically the mean far-infrared luminosities of
  quasars.
  Our quasar compilation at redshifts $0<z<6.5$ and $I$-band
  luminosities $-20<I_{\rm AB}<-32$ is the first to distinguish
  evolution from quasar luminosity dependence in such a study.  We
  carefully cross-calibrate IRAS against Spitzer and ISO, finding
  evidence that IRAS 100$\,\mu$m fluxes at $<1\,$Jy are overestimated
  by $\sim30$\%. We find evidence for a correlation between star
  formation in quasar hosts and the quasar optical luminosities,
  varying as SFR $\propto L_{\rm opt}^{0.44\pm0.07}$ at any fixed
  redshift below $z=2$.  We also find evidence for evolution of the
  mean star formation rate in quasar host galaxies, scaling as
  $(1+z)^{1.6\pm0.3}$ at $z<2$ for any fixed quasar $I$-band absolute
  magnitude fainter than $-28$.  We find no evidence for any
  correlation between star formation rate and black hole mass at
  $0.5<z<4$. Our data are consistent with feedback from black hole
  accretion regulating stellar mass assembly at all redshifts.

\end{abstract}


\section{Introduction}

The interaction between the accretion process into a supermassive
black hole residing at the centre of an active nucleus and star
formation in the host galaxy is fundamental in regulating both galaxy
evolution and the growth of the black hole. In order to understand 
the link between the two processes and assess the possibility of the
two occuring concomitantly, it is important to quantify and constrain
the star formation activity in quasar host galaxies. This is, however, 
a difficult task, as the star formation signature can be diluted by
the strong AGN emission, especially at short (e.g. UV/optical) 
wavelengths. Emission emerging from star formation activity should,
therefore, be looked for in the far-infrared (FIR), where the 
contribution of the AGN (in the form of thermal emission from dust)
should be less important. 

Combined AGN studies with IRAS and ISO already establised the 
presence of strong FIR emission in quasars (see e.g. Verma et al. 2005
and references therein). According to some models, this radiation 
might be explained as the emission of dust distributed in a ``cloud'' 
around the central engine with a $0.5$\,kpc radius (Siebenmorgen et al. 
2004). However, other models estimate dusty tori extending to 
several kpc (e.g. Fritz et al. 2006, Hatziminaoglou et al. 2008)
with additional very large covering factors ($\ge 90\%$). 
Spitzer IRS spectroscopy, revealed in addition to the FIR emission, 
the presence of PAH features in the mid-IR spectra of optically-selected
quasars (Schweitzer et al. 2006), that are difficult to reproduce 
in models assuming dust heated by the hard AGN photons. A number
of other arguments including the likely evaporation of PAH features in
the presence of AGN emission and in the absence of high column
densities (a necessary condition for the AGN to be able to
heat the dust at such large distances) or the simultaneous presence
of star formation evidence at other wavelengths suggest that the 
FIR in quasar host galaxies is more likely to be a tracer of 
star formation. 

Even though Spitzer observations have increased the number of FIR
detections in most low redshift quasars (e.g. Schweitzer et al. 2006)
and their analysis consistently points toward star formation driven
FIR emission, the number of FIR detected quasars is still low. 
%The observed star formation rates of quasars can in principle
%constrain models of the coupled formation of supermassive black holes
%and the stellar mass assembly of their bulges. However, there are few
%direct far-infrared detections of quasars. 
In preparation for the
Herschel ATLAS (Astrophysical Terahertz Large Area Survey), an Open
Time $\sim500\,$deg$^2$ blank-field survey, and in preparation for
targeted Herschel surveys of AGN, we need the best possible estimates
for the quasar fluxes in Herschel bands. Most Sloan Digitized Sky
Survey (SDSS) quasars are not detected individually in the Spitzer
SWIRE Legacy Survey $70\,\mu$m and $160\,\mu$m data
(Lonsdale et al. 2003, 2004) in the SWIRE-SDSS
overlap region (e.g. Hatziminaoglou et al. 2005, 2008).  Submm and
mm-wave observations of $z\simeq2$ and $z>4$ quasars have yielded only
a small number of direct detections (e.g. Omont et al. 1996, Omont et
al. 2001, Carilli et al. 2001, Isaak et al. 2002, Priddey et
al. 2003a, 2003b, Omont et al. 2003, Robson et al. 2004, Beelen et
al. 2006, Petric et al. 2006, Wang et al. 2007). IRAS and ISO detected
just over half of the Palomar-Green quasar sample at $60\,\mu$m
(e.g. Sanders et al. 1989, Haas et al. 2000, 2003). 

In this paper, we
will use stacking analyses to constrain the mean far-infrared luminosities
of quasars, selected over a very wide range in redshift and absolute $I$-band
magnitude. Our quasar compilation spans enough of the $I$-magnitude--redshift
plane to be able to distinguish evolution from quasar luminosity dependence,
which would be impossible in a single $I$-magnitude-limited quasar sample. 
Previous authors have stacked quasar fluxes at one
(usually submm) wavelength, but we will combine a large body of
multi-wavelength far-infrared, submm and mm-wave quasar photometry using
an assumed common SED. We will then show that our conclusions are robust
to reasonable choices of SED. 
We follow SDSS in adopting the concordance cosmology of
$\Omega_{\rm M}=0.3$, $\Omega_\Lambda=0.7$, $H_0=70$\,km/s/Mpc, and in
assuming an optical spectral index of $d\ln S_\nu/d\ln\nu=-0.5$ for
the quasars. In the far-infrared we assume an M82 SED shape from
Efstathiou, Rowan-Robinson \& Siebenmorgen 2000, unless otherwise
stated.

\section[]{Methodology}

\subsection{Sample selection}

\begin{figure*}
  \ForceWidth{10.0cm}
  \hSlide{-4cm}
  \BoxedEPSF{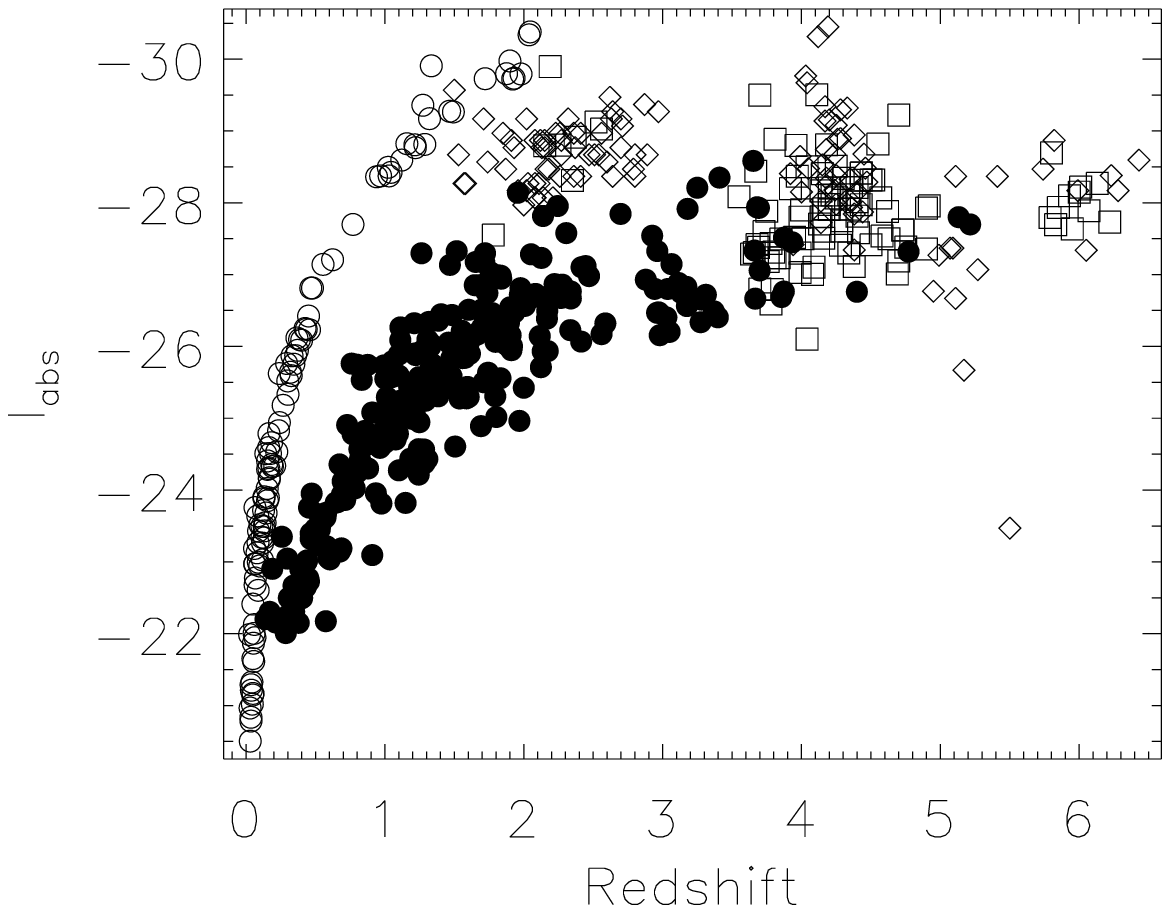}
  \vspace*{-7.2cm}
  \ForceWidth{10.0cm}
  \hSlide{4cm}
  \BoxedEPSF{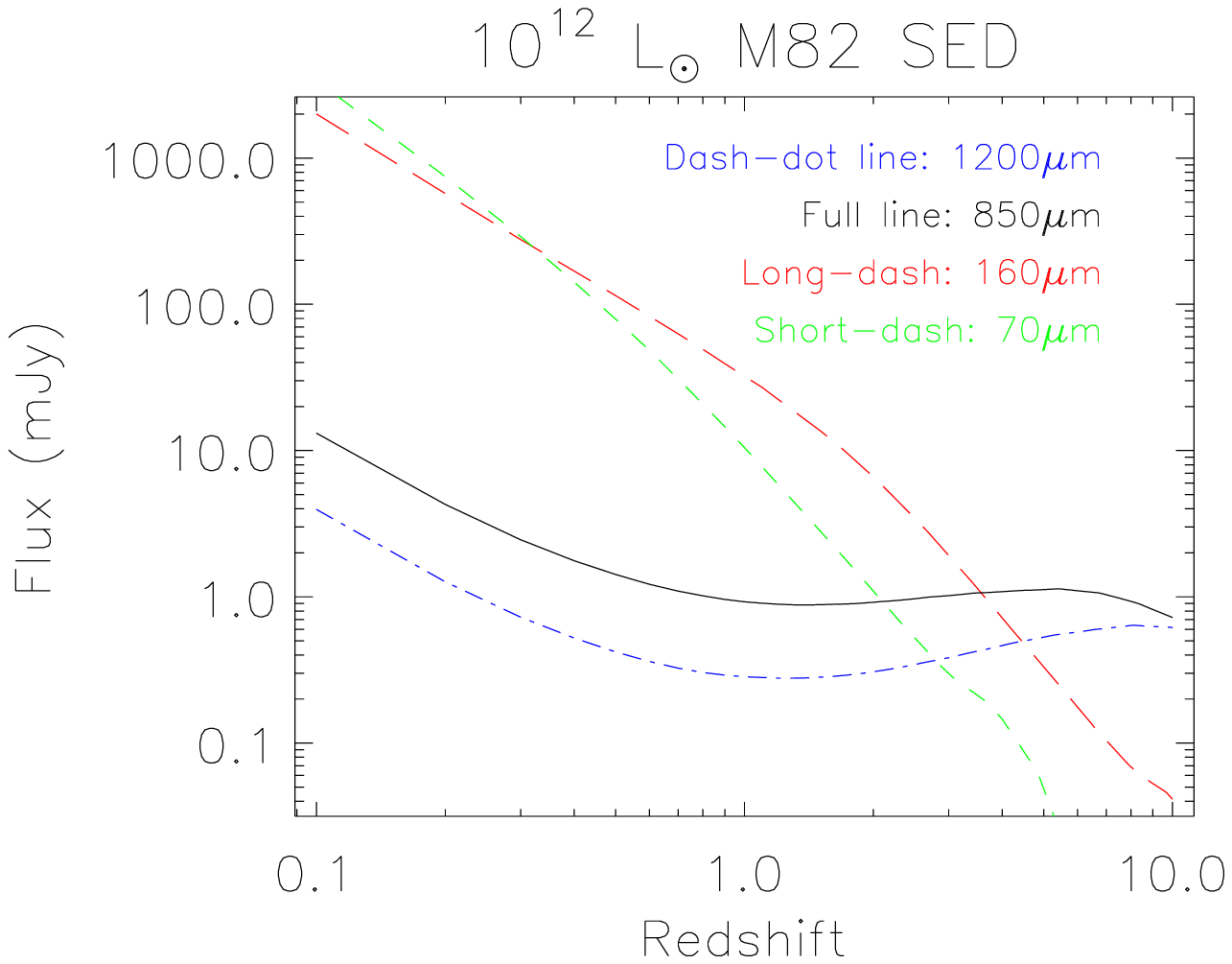}
  \caption{\label{fig:iz} (Left) Absolute I magnitudes of quasars in this
    paper, as a function of redshift. Filled circles represent SDSS
    quasars with SWIRE coverage; open circles represent Palomar-Green
    quasars with IRAS and B-band data; diamonds represent quasars
    observed at 850$\,\mu$m; open squares represent quasars observed at
    1200$\,\mu$m (see text for details). 
    Note that adding the SDSS quasars to the other samples
    greatly improves the coverage of the optical luminosity - redshift
    plane. (Right) Predictions for the far-infrared and submm fluxes
    of an M82 SED normalised to $\nu L_\nu=10^{12}L_\odot$ at $100\,\mu$m. 
    Note that the stacked signal
    from SDSS quasars at $z\simeq2$ are well-matched in far-infrared luminosity
    to the expected stacked signal from Palomar-Green quasars at
    $z\simeq0.5$.}
\end{figure*}

Figure \ref{fig:iz} shows the absolute I magnitudes of the SDSS
quasars with SWIRE $70\,\mu$m and/or $160\,\mu$m coverage, against
redshift. The SWIRE data was retrieved on 17th August 2007 and
comprises version 2 products in the ELAIS N1 and ELAIS N2 fields, and
version 3 products in the Lockman Hole field. There are 281 DR5 SDSS
quasars (Adelman-McCarthy et al. 2007)
in the SWIRE fields with 70$\,\mu$m and/or $160\,\mu$m
coverage. Of these, 264 have 70$\,\mu$m coverage and 261 have 160$\,\mu$m
coverage. ELAIS N1 is only partly covered by SDSS. 
There is no SDSS data for the Southern SWIRE fields XMM-LSS, ELAIS S1 or
CDF-S, though this data was retrieved for calibration (see below); the
version numbers are 2, 3 and 3 respectively. Figure \ref{fig:iz} also shows for
comparison the Palomar-Green sample with far-infrared photometry from
IRAS and ISO (Sanders et al. 1989, Haas et al. 2000, 2003), in which
non-detections have been re-measured using the SCANPI IRAS fluxes 
discussed below. Figure \ref{fig:iz} also shows a compilation of the quasars
observed at 850$\,\mu$m and 1200$\,\mu$m (Omont et al. 1996, Omont et
al. 2001, Carilli et al. 2001, Isaak et al. 2002, Omont et al. 2003,
Priddey et al. 2003a, 2003b, Robson et al. 2004, Wang et
al. 2007). The addition of the SDSS quasars to these data sets greatly
improves the coverage of the optical luminosity-redshift plane. This
will make it possible to make the first constraints on the evolution
and luminosity dependence of star formation in quasar hosts. The
right-hand panel of figure \ref{fig:iz} demonstrates the far-infrared
luminosities probed by the various multi-wavelength data sets. 

\subsection{SWIRE photometry}

The SWIRE 70$\,\mu$m and 160$\,\mu$m images are supplied calibrated to surface
brightness units of MJy/sr. We intend to measure the total mean point
source flux of our targets, rather than the mean surface brightness at
their locations, so we need to convert the units. We first note that a
Gaussian point source with a peak flux of unity will have a total flux
of $F=\pi\theta^2_{\rm FWHM}/(4\ln2)$, where $\theta_{\rm FWHM}$ is
the full-width half maximum of the point spread function. If 
$\theta_{\rm FWHM}$ is
measured in arcseconds, then this also gives the conversion between
total point source flux and the flux per square arcsecond at the peak.

\begin{figure*}
  \ForceWidth{10.0cm}
  \hSlide{-4cm}
  \BoxedEPSF{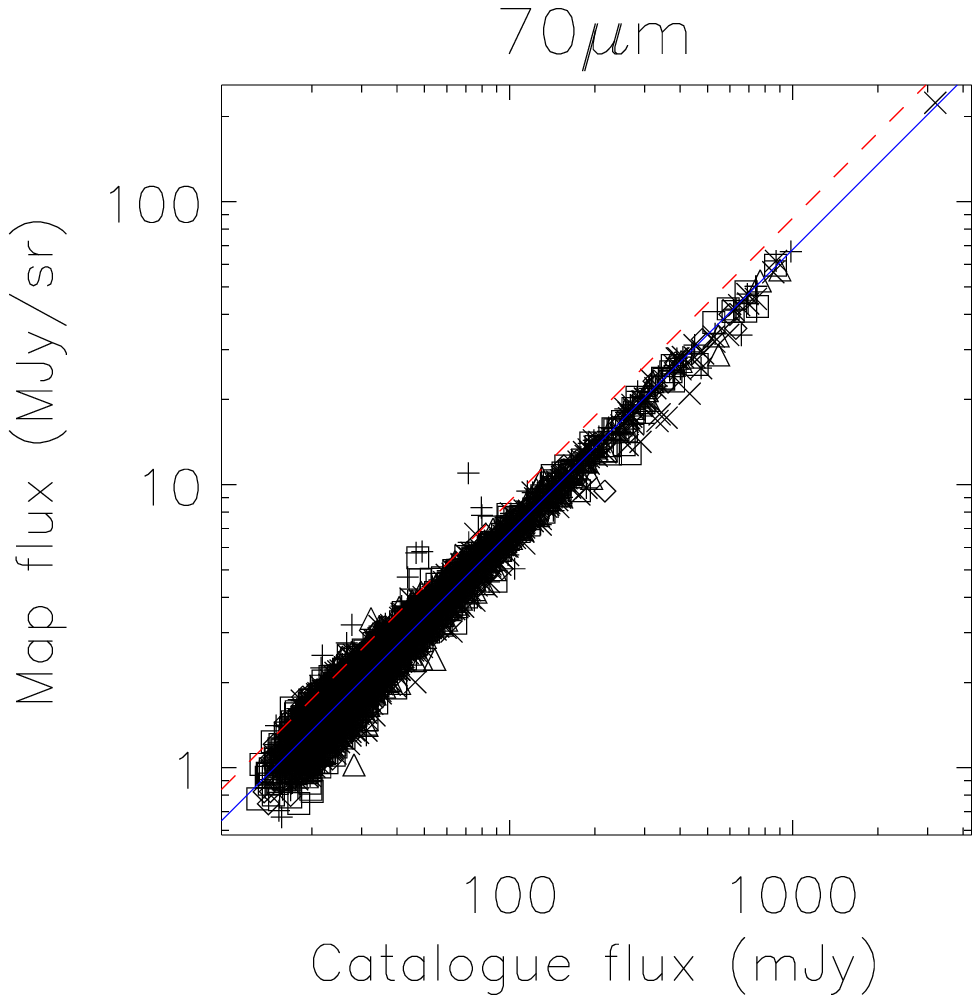}
  \vspace*{-7.2cm}
  \ForceWidth{10.0cm}
  \hSlide{4cm}
  \BoxedEPSF{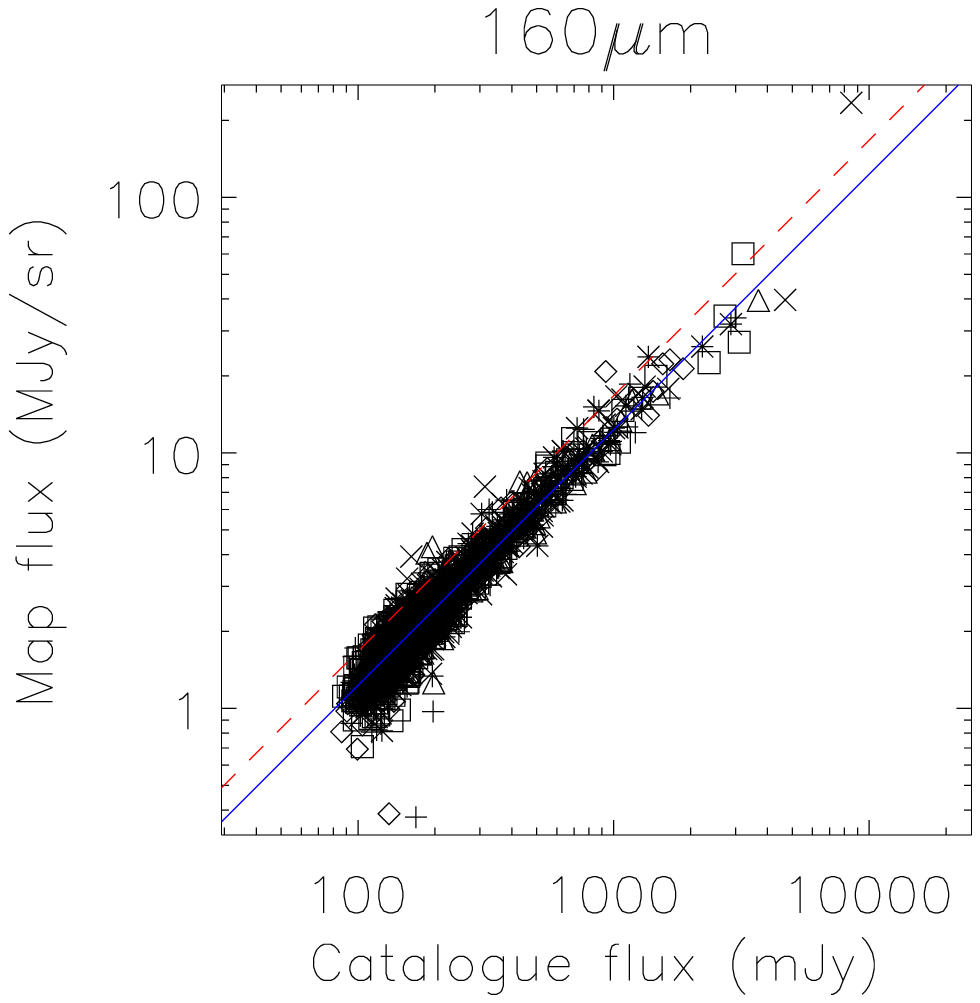}
  \caption{\label{fig:swire_calibration}Conversion between map surface
    brightness units and point source total fluxes, using the SWIRE
    catalogues (not restricted to SWIRE quasars). The 70$\,\mu$m data
    are shown to the left, and the 160$\,\mu$m data to the right. The
    dashed (red) line shows the predicted conversion discussed in the text, and
    the full (blue) line gives the empirical conversion. Numerical values for
    the conversions are given in table \ref{tab:conversions}. Symbols:
    + ELAIS N1, {\tt *} ELAIS N2, $\diamond$ CDFS, $\triangle$ ELAIS
    S1, $\Box$ Lockman Hole, $\times$ XMM-LSS.}
\end{figure*}

Using $\theta_{\rm FWHM}=1.22\lambda/D$, where $\lambda$ is the
observed wavelength and $D=85\,$cm is the diameter of the telescope's
primary mirror, we predicted the conversion between point source flux
and surface brightness. This is shown in table
\ref{tab:conversions}. We also derived an empirical conversion
obtained by comparing the SWIRE point source catalogue fluxes with the
background-subtracted map fluxes measured at the positions of the
SWIRE sources. This comparison is shown in figure
\ref{fig:swire_calibration}. The empirical conversion is some 29-36\%
higher than the theoretical prediction, which may be due to the finite
size of the map pixels, or a non-Gaussian shape to the point spread
function shape (e.g. more resembling an Airy function). We adopt the
empirical conversion in the analysis below.

\begin{table}
\begin{tabular}{lll}
 & Predicted conversion & Emprical conversion\\
$70\,\mu$m & 11.44 & 14.76 \\
$160\,\mu$m & 59.76 & 81.14 \\
\end{tabular}
\caption{\label{tab:conversions}Point source fluxes in mJy for a
  source in the SWIRE maps with a central surface brightness of 1\,MJy/sr.}
\end{table}

Although the maps are supplied with foreground DC offsets removed, we
opted to subtract the mean flux level from each image. This ensures
that the average total flux contribution is identically zero from each
point source not associated with the SDSS quasars. The stacking
methodology is to measure the point source fluxes from the SWIRE
images at the positions of the SDSS quasars, and search for a
significant deviation from zero flux. Photometric errors were
estimated by measuring the standard deviation of the SWIRE images. The
SDSS quasar astrometric errors are negligible compared to the SWIRE
70-160$\,\mu$m beams.

\subsection{IRAS photometry}
SCANPI IRAS flux estimates have recently been adapted to report
negative AMP flux estimates, where acceptable fits are available. This
would make this estimator suitable to stacking analyses, but we found
positive fluxes on average reported at positions randomly offset from
our targets (as noted also by e.g. Morel et al. 2001). This was found
to be due to the SCANPI algorithm allowing the source position to
vary, so that the fits tended to gravitate to nearby noise features
when the signal-to-noise is very low. Therefore, we re-fit the coadded
SCANPI profiles allowing only the point source flux to vary, using the
appropriate point source response function for the predominant scan
direction (A. Alexov, priv. comm.). We used the background-subtracted
median coadded scans. At 60$\,\mu$m, we subtracted an additional background
estimate obtained by fitting a Gaussian to the off-source data
histogram for each coadded scan (where ``off-source'' is defined as
where the template is $<1$\% of its maximum value). At 100$\,\mu$m, the
stronger baseline drifts visible in the coadded scans suggested a more
local background subtraction. We defined the 100$\,\mu$m on-source width to
be where the template is $>0.1$\% of its maximum, and subtracted the
mean of the coadded scan data in one template-width either side of the
target. Note that our flux calibrations discussed below were found to
depend weakly on the sky subtraction algorithm, but were not
independent of it.

We tested the flux calibration of stacked fluxes using 70$\,\mu$m sources
selected from the Spitzer legacy surveys SWIRE and Formation and
Evolution of Planetary Systems (FEPS, e.g. Meyer et al. 2004,
Hillenbrand et al. 2008), and using 90$\,\mu$m sources selected from the
European Large Area ISO Survey (ELAIS, Rowan-Robinson et
al. 2004). Note that the SWIRE and ELAIS surveys were conducted in
regions of low cirrus; the FEPS survey, while having fewer sources, is
more widely-distributed. We selected all SWIRE 70$\,\mu$m sources in the
flux range 12-300mJy, then starting from the brightest, we rejected
any source closer than 30' to any selected source. This ensured that
both our source fluxes and their background estimates are
statistically independent.  We adopted the same selection procedure
for ELAIS. For FEPS, we selected all sources with 70$\,\mu$m detections at
$5\sigma$ or above. For FEPS, SWIRE and ELAIS sources, we extracted IRAS
fluxes in a $3\times3$ grid centred on the target, with grid positions
separated by 20'.

\begin{figure*}
  \ForceWidth{10.0cm}
  \hSlide{-4cm}
  \BoxedEPSF{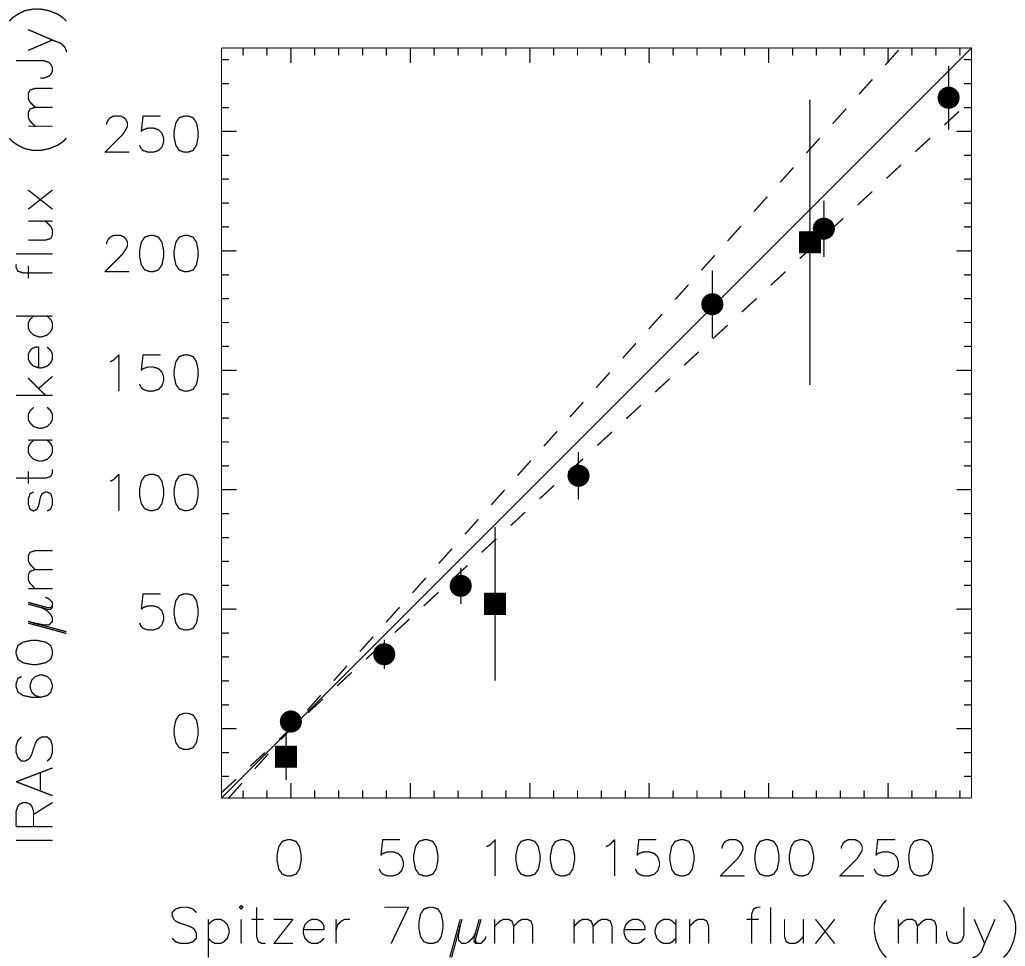}
  \vspace*{-7.2cm}
  \ForceWidth{10.0cm}
  \hSlide{4cm}
  \BoxedEPSF{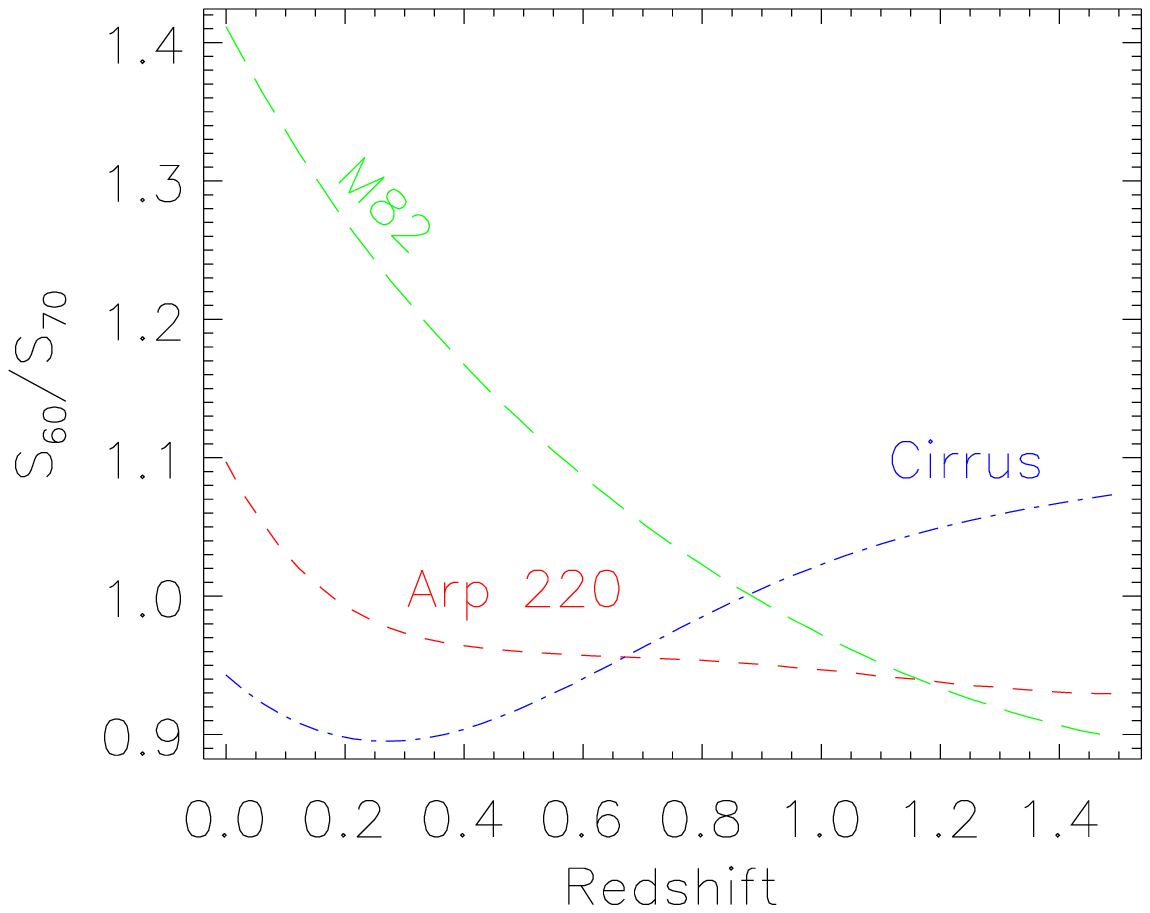}
  \caption{\label{fig:60um_calibration}The left panel shows the
    unweighted mean IRAS 60$\,\mu$m AMP SCANPI fluxes for Spitzer
    70$\,\mu$m SWIRE sources (circles) and FEPS sources (squares). The
    errors are the usual $\sigma/\sqrt{N}$ estimate of the noise on the
    mean, except for the zero-flux bins where the signal and noise are
    estimated from a Gaussian fit to the measurements as discussed in
    the text. The full line shows the 1:1 relation, and the dashed
    lines show the expected range of colours at $0.5<z<1.5$ for M82,
    Arp 220 and a cirrus spectrum, from Efstathiou et
    al. 2000. Colours were calculated using the respective
    wavelength-dependent system responses, following the prescription
    in {\tt
      http://spider.ipac.caltech.edu/staff/lord/MIPS/MIPS.html}. The
    redshift dependences of this colour for these SEDs are shown in
    the right panel, with Arp 220 as short dashes, M82 as long dashes,
    and cirrus as dash-dot.}
\end{figure*}

\begin{figure*}
  \ForceWidth{10.0cm}
  \hSlide{-4cm}
  \BoxedEPSF{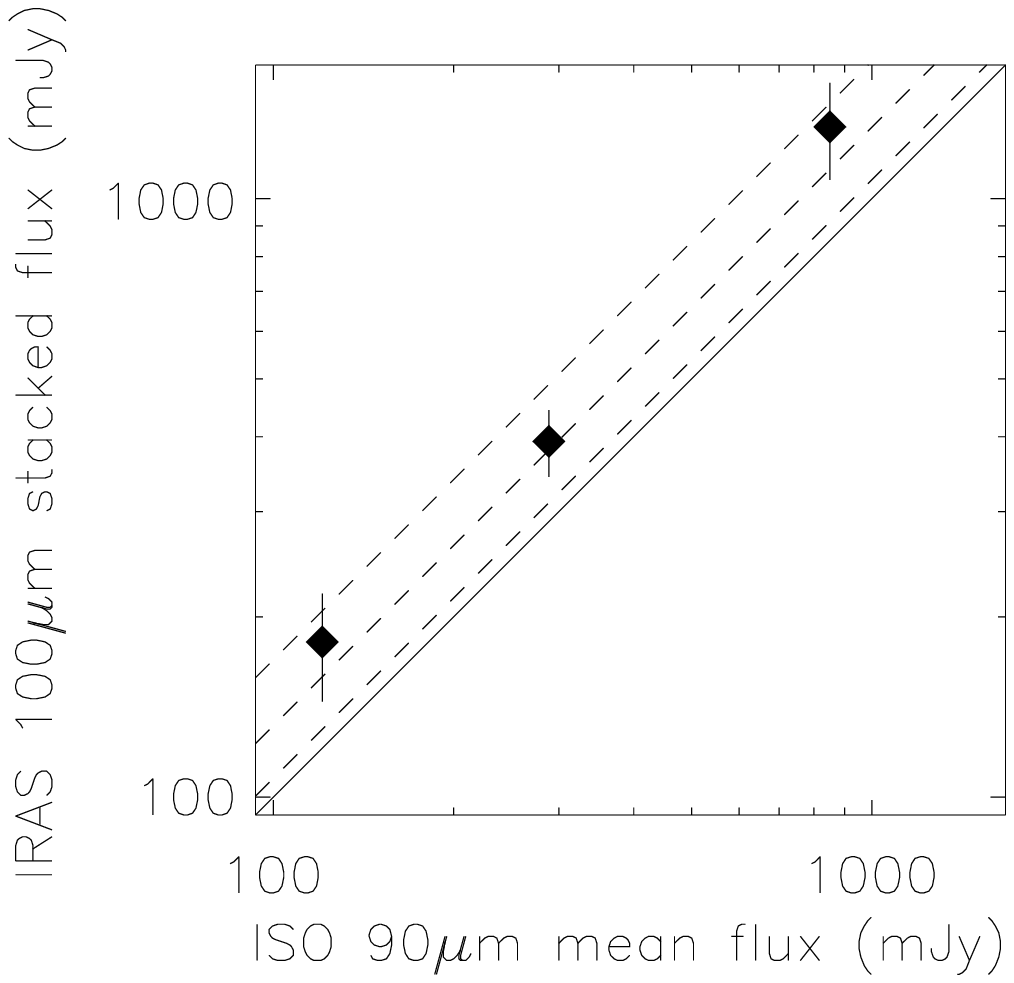}
  \vspace*{-7.2cm}
  \ForceWidth{10.0cm}
  \hSlide{4cm}
  \BoxedEPSF{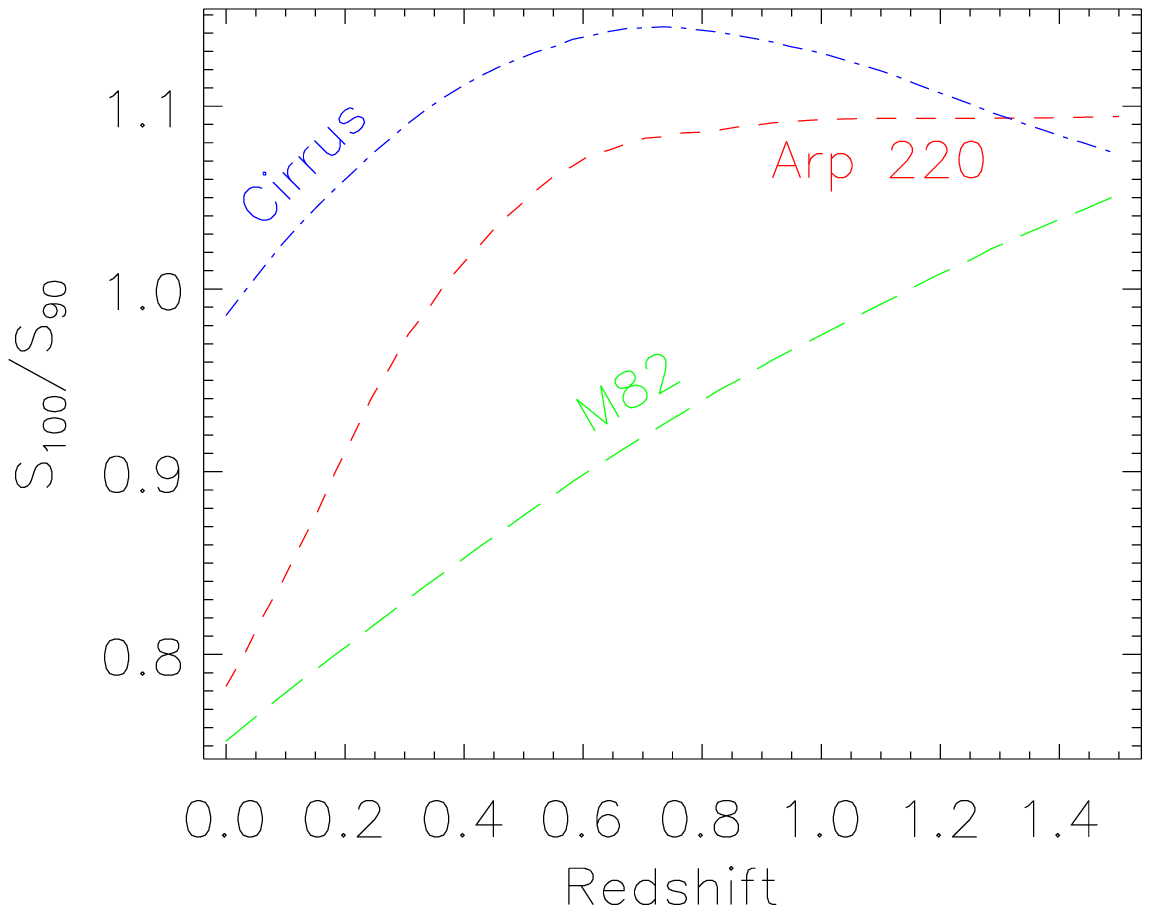}
  \caption{\label{fig:100um_calibration}The left panel shows the
    unweighted mean IRAS 100$\,\mu$m stacked fluxes, as a function of
    ELAIS ISO 90$\,\mu$m fluxes. Errors are as in figure
    \ref{fig:60um_calibration}. The dashed lines show the mean
    IRAS-ISO calibration offset found by H\'{e}raudeau et al. (2002),
    and the RMS of their sample. The right panel shows the expected
    IRAS-ISO colours expected for the SEDs in figure
    \ref{fig:60um_calibration}.}
\end{figure*}

To estimate the noise in each SCANPI flux estimate, we tried two
approaches: firstly, estimating the variance in the coadded scan from
Gaussian fits to the data histogram, then propagating the noise in the
best-fit point source amplitude assuming the data points are
statistically independent; secondly, measuring the variance of the
point source amplitude estimates in the eight offset sky
positions. The latter should give signal-to-noise histograms with zero
mean and unit variance, while for the former the histogram of $(S_{\rm
  IRAS}-S_{\rm Spitzer})/N_{\rm IRAS}$ (where $S$ and $N$ are signal
and noise respectively) should also have zero mean and unit
variance. The SCANPI data points are not statistically independent,
and we found that propagating the noise led to error estimates too
small by a factor of around 3.6 at 60$\,\mu$m, and 5.7 at
100$\,\mu$m. The offset positions had consistent noise estimates, but
have the disadvantage that the estimates are not local to the
target. A visual inspection of the IRAS maps around our Palomar-Green
targets (A. Alexov, priv. comm.) suggested problematic cirrus
structure near several targets. We therefore adopted the propagated
noise estimate, scaled by a factor of 3.7 (5.7) at 60$\,\mu$m
(100$\,\mu$m) to account for correlations between the SCANPI data
points. Cirrus structure can in principle be alleviated with a matched
filter tuned to the power spectrum of the background (e.g. Vio et
al. 2002), though in this case the IRAS detector responsivity
variations would present a significant complication.

We found the SCANPI 1002 coadded scans (median combination of IRAS
scans) to give the best signal-to-noise. Figures
\ref{fig:60um_calibration} and \ref{fig:100um_calibration} show the
unweighted mean average IRAS fluxes of the fainter Spitzer and ISO
sources in broad flux bins. This stacking methodology appears to give
unbiased estimates even at the faintest 70$\,\mu$m fluxes tested,
e.g. $\sim40\,$mJy at 70$\,\mu$m. This is much fainter than the mean
60$\,\mu$m fluxes of Palomar-Green quasars, which we estimate to be
$201\pm35\,$mJy. For the offset positions, we estimated the signal and
noise from a Gaussian fit to the histogram of the measurements, in
order to avoid serendipitous sources. The stacked 60$\,\mu$m fluxes at
random offset positions were consistent with zero
(e.g. $3.0\pm1.5$\,mJy for SWIRE, see figure
\ref{fig:60um_calibration}).

At 100$\,\mu$m, the stacked flux at blank-field positions offset from
SWIRE sources was $-6.7\pm6.3\,$mJy. However, we found evidence for
flux calibration discrepancy between ELAIS and our 100$\,\mu$m stacked
fluxes (figure \ref{fig:100um_calibration}).  H\'{e}raudeau et
al. 2002 found that the ELAIS 90$\,\mu$m point source fluxes were
lower than IRAS FSC 100$\,\mu$m fluxes by a factor $0.76\pm0.17$,
after taking into account the colour corrections. They attributed this
to systematic IRAS overestimates in the faintest catalogues fluxes,
citing Moshir et al. 1992; however, we find no evidence for such a
calibration offset in our 60$\,\mu$m IRAS stacks. We compared the
ELAIS and SWIRE flux calibrations by interpolating the 70$\,\mu$m and
160$\,\mu$m fluxes to obtain 90$\,\mu$m flux estimates, and found a
1:1 correlation with the ELAIS fluxes, so we think it unlikely the
ELAIS flux calibration is at fault. We also calculated the
100$\,\mu$m:90$\,\mu$m colour for our three SED models (figure
\ref{fig:100um_calibration}), and found this too small an effect to
account for the discrepancy. A similar discrepancy has been noted by
Jeong et al. (2007), comparing IRAS 100$\,\mu$m and AKARI 90$\,\mu$m
fluxes. We have therefore chosen to adopt the H\'{e}raudeau et
al. correction factor of 0.76 to our stacked 100$\,\mu$m IRAS
fluxes. In the appendix, we present our SCANPI photometry of PG
quasars and discuss the problematic cases.

\section{Results}
\subsection{Stacked flux results}
\begin{figure*}
  \ForceWidth{10.0cm}
  \hSlide{-4cm}
  \BoxedEPSF{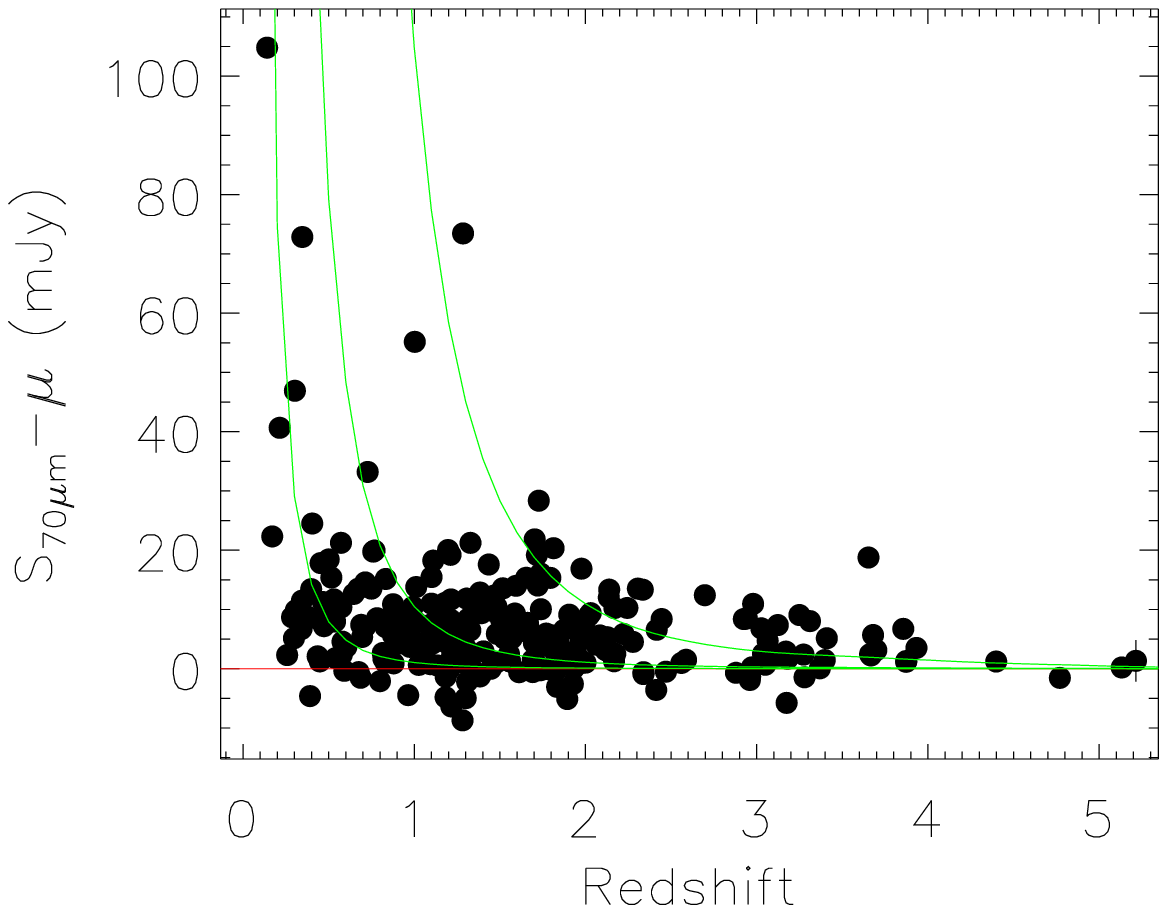}
  \vspace*{-7.2cm}
  \ForceWidth{10.0cm}
  \hSlide{4cm}
  \BoxedEPSF{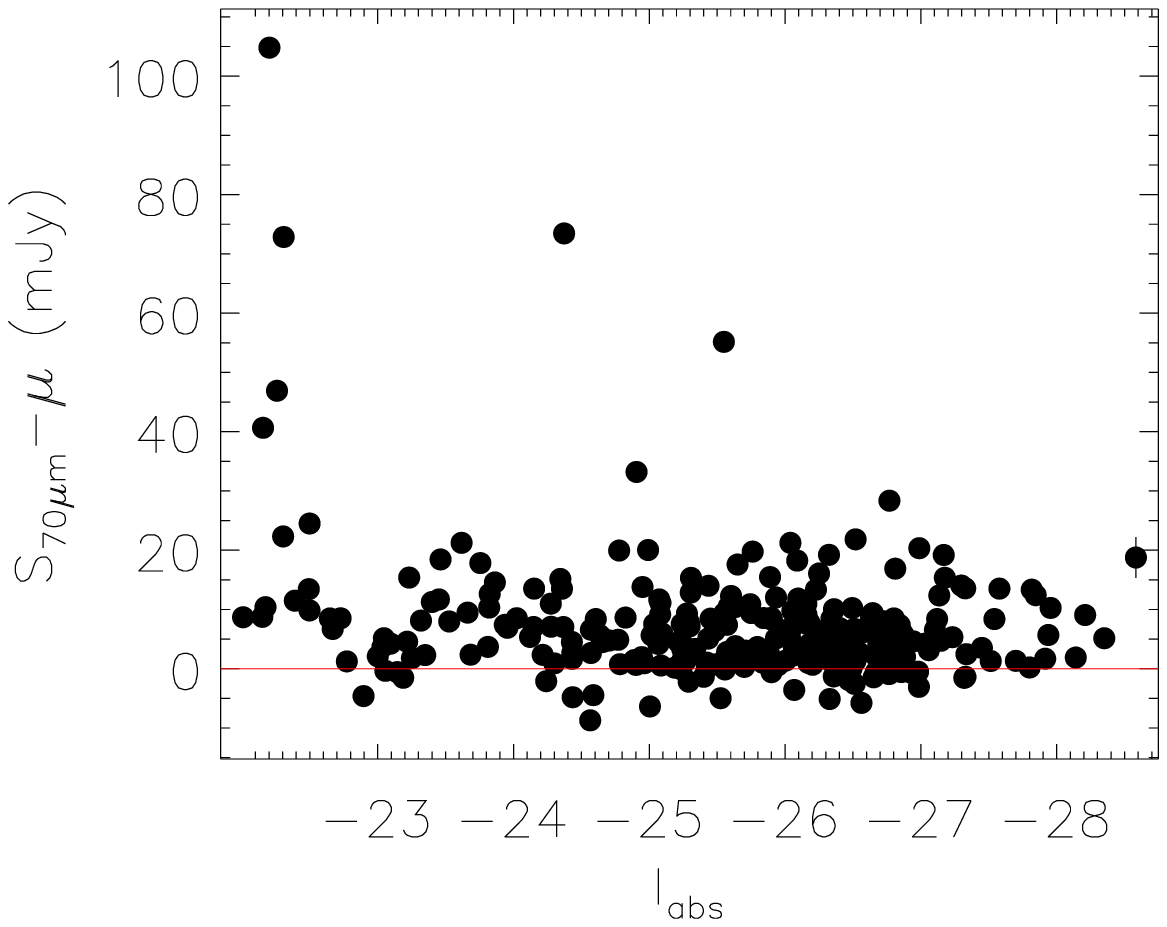}
  \vspace*{-1cm}
  \ForceWidth{10.0cm}
  \hSlide{-4cm}
  \BoxedEPSF{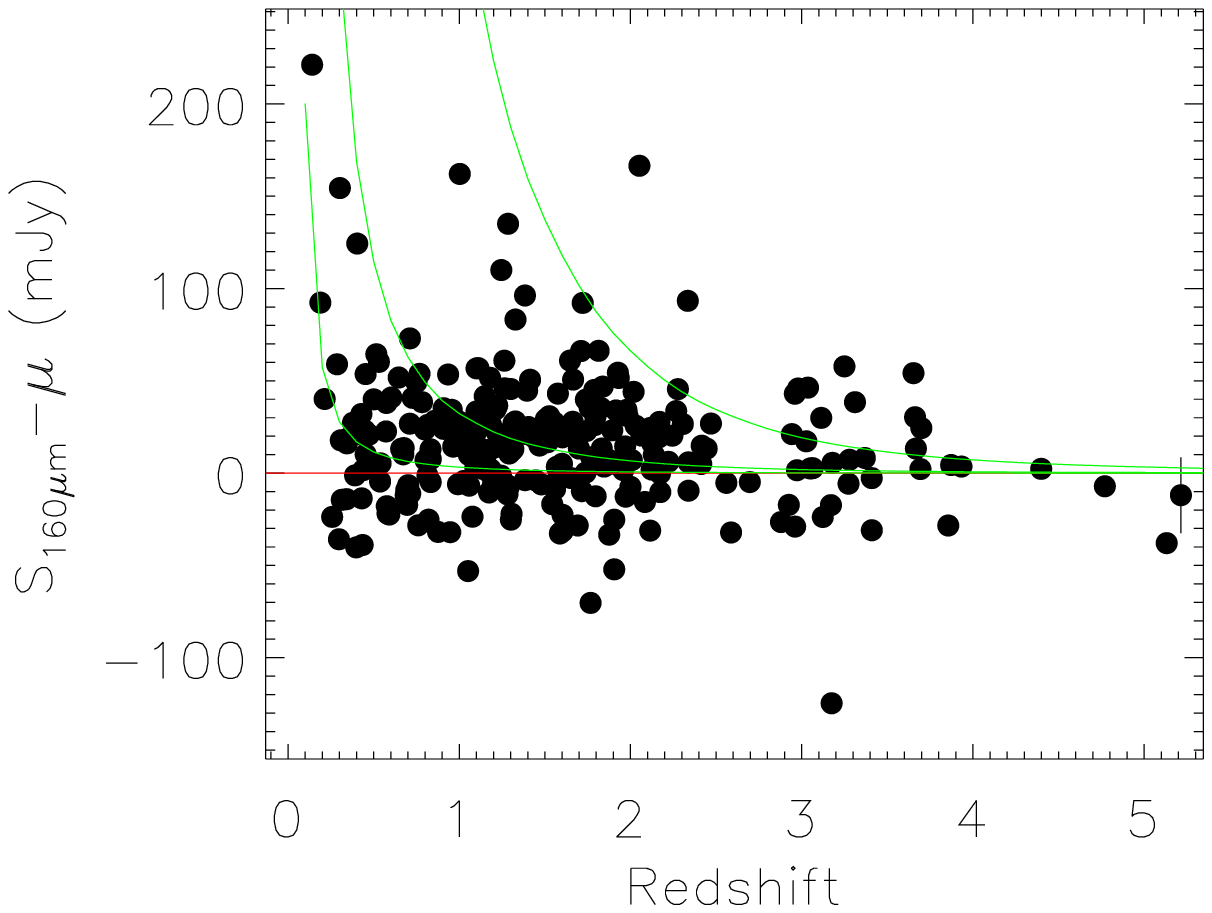}
  \vspace*{-7.2cm}
  \ForceWidth{10.0cm}
  \hSlide{4cm}
  \BoxedEPSF{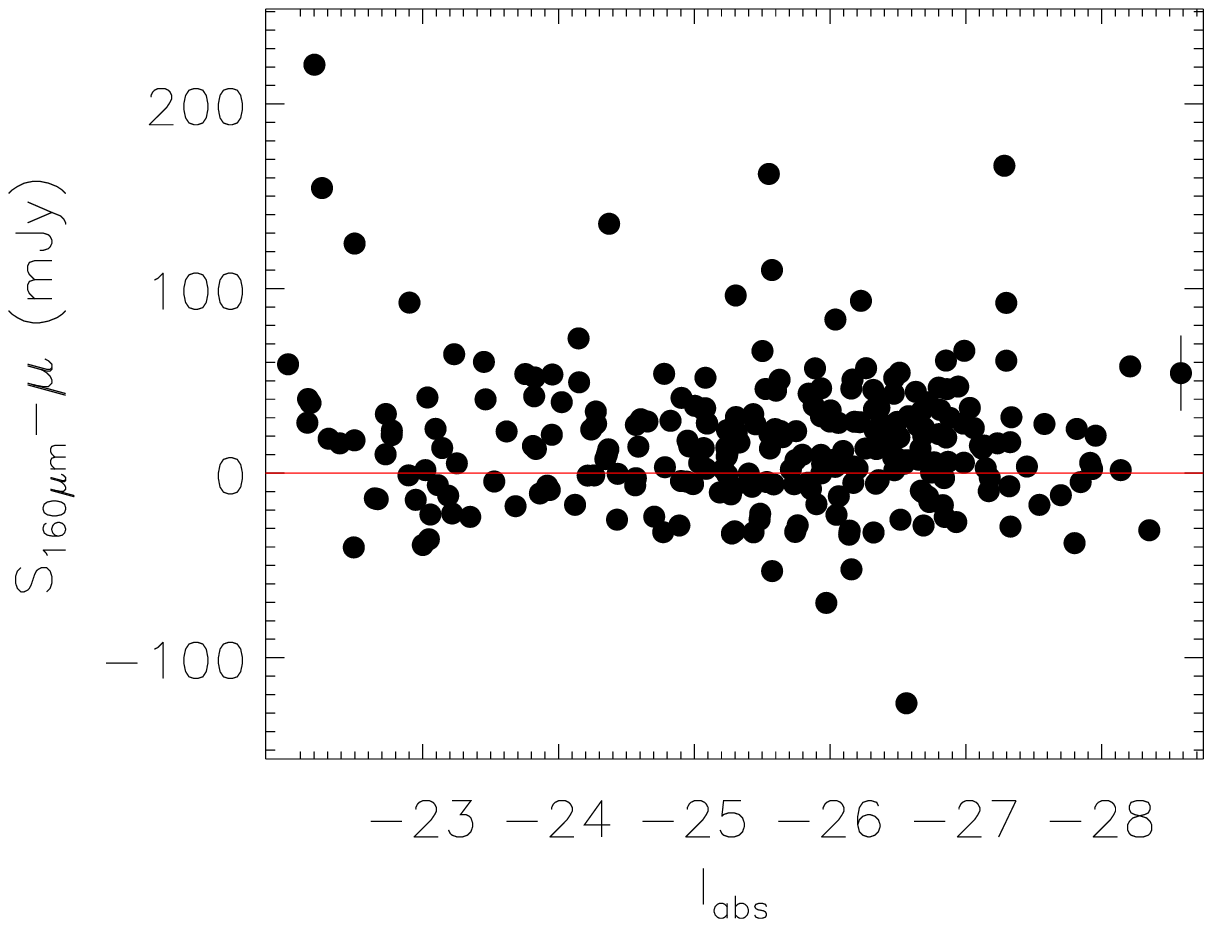}
  \caption{\label{fig:sdss_swire} SWIRE 70$\,\mu$m and 160$\,\mu$m
    fluxes at the positions of SDSS quasars, as a function of the
    quasar absolute magnitudes and redshifts. A background sky level
    has been subtracted, as discussed in the text. The red line shows
    the prediction for zero flux, while the green curves show the
    location of an M82 starburst SED normalised to $\nu L_\nu=10^{11}$,
    $10^{12}$ and $10^{13}L_\odot$ at $100\,\mu$m. 
    Note that at redshifts $z<3$, and at
    most absolute magnitudes, the quasars appear significantly above
    the zero flux line. Error bars have been suppressed for clarity,
    except for a single representative error bar in each panel. Flux
    errors were estimated from Gaussian fits to the pixel histograms.}
\end{figure*}

Figure \ref{fig:sdss_swire} shows the SWIRE 70-160$\,\mu$m fluxes for
the SDSS quasars as a function of redshift and absolute
magnitude. Error bars have been suppressed for clarity, except for a
single representative example in each panel. The SDSS quasar
population appears bimodal or at least with a significant skew to 
bright fluxes for a minority of objects, with a small number of far-IR-loud
objects at both 70$\,\mu$m and 160$\,\mu$m.  For the bulk of the
population, there also appears to be a significant positive flux at
the positions of the SDSS quasars at redshifts $z<3$, and at most
absolute magnitudes. Of the quasars at $0.5<z<1.5$, 6/114 have
70$\,\mu$m fluxes above $20\,$mJy, and 6/113 have 160$\,\mu$m fluxes
above $70\,$mJy. In the $1.5<z<2.5$ redshift bin, 3/86 have 70$\,\mu$m
fluxes above $20\,$mJy, and 3/84 have 160$\,\mu$m fluxes above
$70\,$mJy. The green curves show the predicted flux for an M82 SED,
normalised to $10^{11}L_\odot$, $10^{12}L_\odot$ and $10^{13}L_\odot$.

\begin{figure*}
  \ForceWidth{10.0cm}
  \hSlide{-5cm}
  \BoxedEPSF{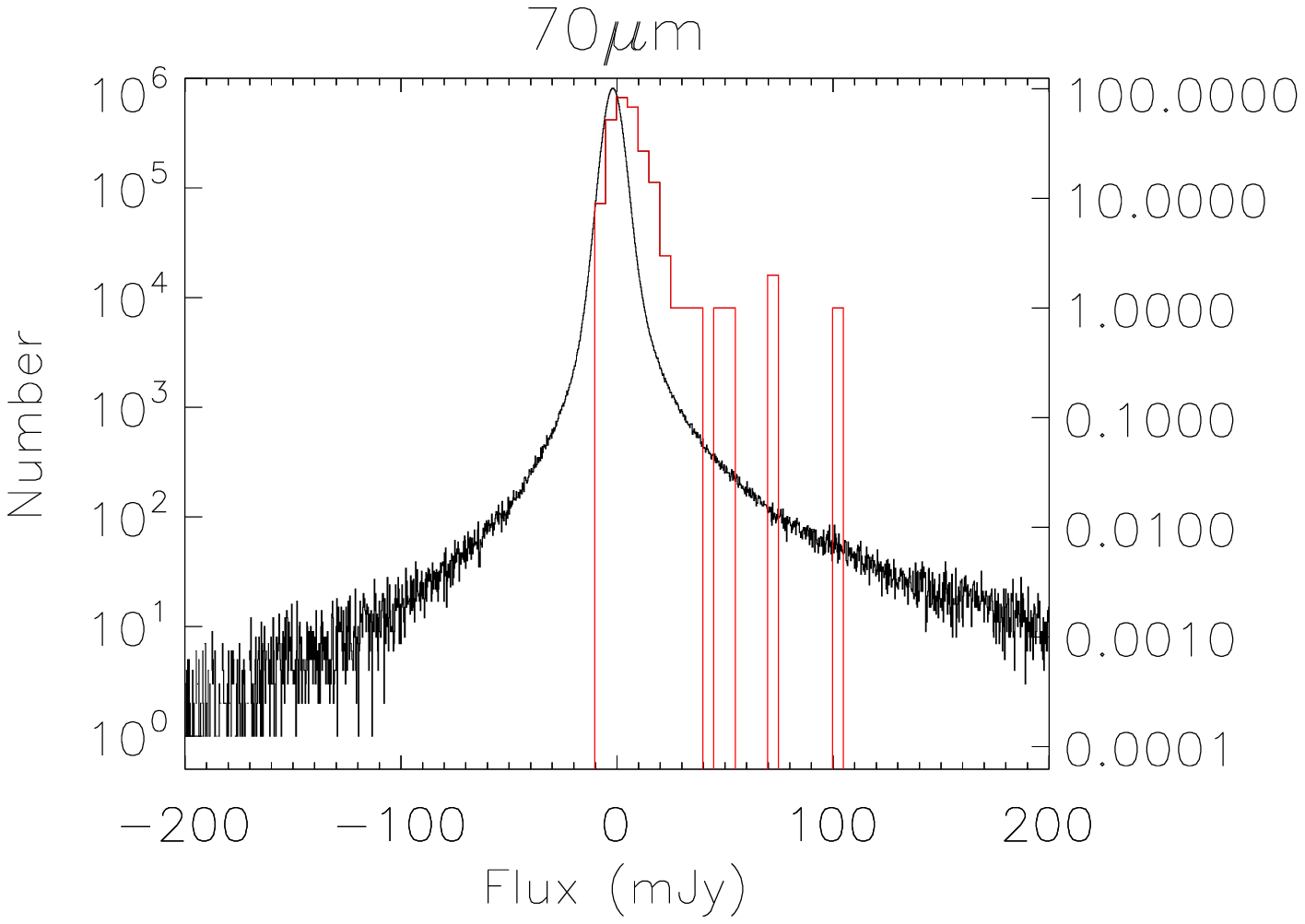}
  \vspace*{-7.2cm}
  \ForceWidth{10.0cm}
  \hSlide{5cm}
  \BoxedEPSF{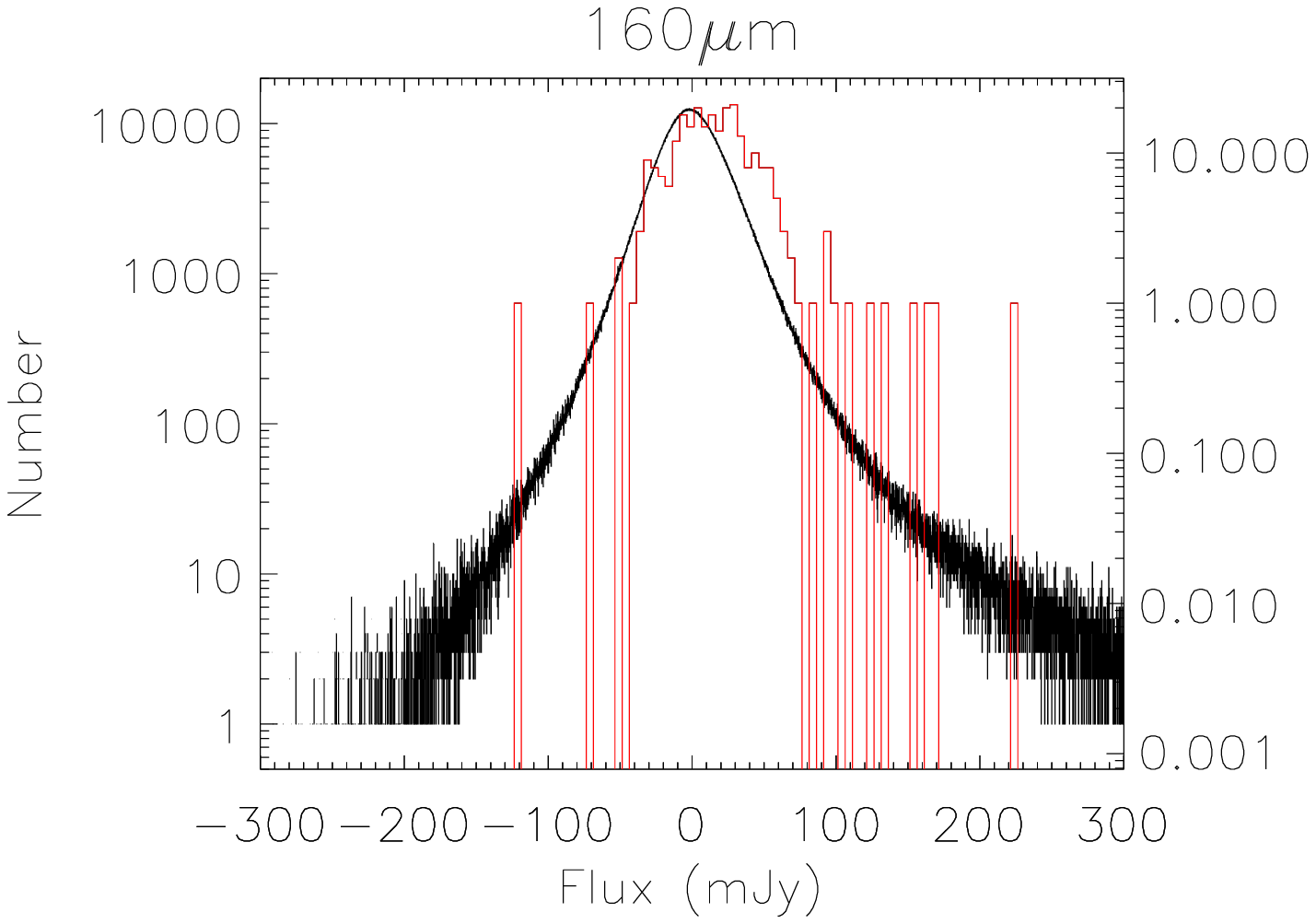}
  \caption{\label{fig:histograms}Histogram of map fluxes (fine bins, black line,
    left-hand ordinates) compared to the fluxes at the positions of
    the SDSS quasars (broad bins, red line, right-hand ordinates). Note the clear
    excess of positive fluxes at the positions of the SDSS quasars,
    i.e. the red lines are higher than the black lines on the right
    hand sides.}
\end{figure*}

To test whether the skewed distribution is caused by a
distinct population of far-IR-loud quasars, or whether unrelated
far-IR-bright foreground companion galaxies are responsible, we
compared the fluxes at the positions of SDSS quasars with those in the
map as a whole, following the methodology of Serjeant et
al. (2004). This test is preferable to Monte Carlo randomizations of
the quasar positions, since it compares the quasar fluxes with the
entire distribution of map fluxes, rather than a randomly-selected
subset. The results are shown in figure \ref{fig:histograms}.  The
observed skewness in the quasar far-infrared fluxes is therefore not
due to a pathological distribution of map fluxes. In table
\ref{tab:swire_stacks}, we present the average far-infrared fluxes for
the SDSS quasars in redshift bins, both with and without the high-flux
population.

\begin{table*}
\begin{tabular}{l || l | l | l | l | l | l} 
Wave- & $0.5<z<1.5$ & $0.5<z<1.5$           & $1.5<z<2.5$ & $1.5<z<2.5$           & $0.5<z<2.5$ & $0.5<z<2.5$\\
length &             & no far-IR bright QSOs &             & no far-IR-bright QSOs &             & no far-IR bright QSOs\\
\hline
70$\,\mu$m & 7.64$\pm$0.95 & 5.98$\pm$0.56 & 6.11$\pm$0.65 & 5.48$\pm$0.56 & 6.98$\pm$0.61 & 5.77$\pm$0.40\\
160$\,\mu$m & 19.3$\pm$3.0 & 14.2$\pm$2.3 & 16.1$\pm$3.6 & 12.3$\pm$3.0 & 17.9$\pm$2.3 & 13.4$\pm$1.8\\
\end{tabular}
\caption{\label{tab:swire_stacks}
  Mean far-infrared fluxes in mJy for SDSS quasars in the indicated
  redshift intervals. The errors quoted are the usual error on the
  mean, $\sigma/\sqrt{N}$, where $\sigma$ is the sample standard
  deviation and $N$ is the sample size. Quasars are deemed to be
  far-infrared bright in the 70$\,\mu$m stack if they have a flux of
  over 20\,mJy. In the 160$\,\mu$m stack, the corresponding threshold
  is 70\,mJy. Sky background levels have been subtracted as discussed
  in the text.} 
\end{table*}

It is clear from the curves in 
figure \ref{fig:sdss_swire} that although the SDSS
quasars span a narrow range of far-infrared fluxes, they span a wide
range of far-infrared luminosities. The right-hand panel in figure
\ref{fig:iz} demonstrates that the stacked 160$\,\mu$m signal from
$z\simeq2$ quasars in SDSS covers comparable far-infrared luminosities
to the expected stacked signal (a few times fainter than the IRAS PSC)
from $z\simeq0.5$ Palomar-Green quasars, for the assumed M82 SED.
Also, it is clear that combining the Palomar-Green, SDSS and SCUBA
quasars gives enough coverage of the optical luminosity-redshift plane
to remove the degeneracy between trends in luminosity and in redshift
(essentially Malmquist bias in its original sense, Malmquist 1924; see
also Teerikorpi 1984).

\subsection{Stacked luminosity results}\label{sec:stacked_luminosity_results}
It is not obvious what the best stacking statistic is where a large
number of the sample have high signal-to-noise direct detections, as
is the case with our quasars. Variance weighting the stacks would
result in high signal-to-noise measurements dominating, but these
measurements need not be in agreement with each other, because the
population has an intrinsic dispersion. For example, if one has two
quasars with fluxes $100\pm1$mJy and $0\pm1$mJy, what can one say
about the average in this population? Clearly, it would not be right
to quote $50\pm0.7$mJy for this average on the evidence of those two
quasars.

We have opted to regard flux measurements of individual quasars as
attempts to measure the mean of the population, so the dispersion in
the population is a noise term on these measurements. The noise on any
particular estimate of the population mean is therefore the quadrature
sum of the flux error and the population dispersion.

But what is an appropriate value for this population dispersion? We
have opted to determine this from our data simultaneously with the
population mean. If $x_i$ are our measurements, each with a
measurement noise level $\sigma_i$, then our data should have the
following distribution:
\begin{equation}
\rho(x,\sigma)=\frac{1}{\sqrt{2\pi(\sigma^2+\sigma_0^2)}} 
               \exp\left (-\frac{(x-\mu)^2}{2(\sigma^2+\sigma_0^2)}\right )
\label{eqn:rho}
\end{equation}
where $\mu$ is the population mean and $\sigma_0$ is the population
dispersion. We find the maximum-likelihood solutions for $\mu$ and
$\sigma_0$ from our data by maximising $\Pi\rho(x_i,\sigma_i)$, and
estimate the parameters' 68\% confidence bounds directly from the
likelihood surface. Note that there is no covariance between these
parameters; this follows from the fact that the expected values of
measurements are independent of their noise levels. We used numerical
simulations to verify that our maximum-likelihood solutions using this
procedure are unbiased estimators of the underlying values, and to
verify our confidence bounds.

This estimator worked well where we had more than three objects in a
bin, but encountered numerical problems with some bins containing only
two or three objects. There is also not enough information to
constrain both parameters when there is only a single object. One
option is to neglect the population dispersion $\sigma_0$ in these
problematic cases, but this would lead to over-optimistic error
estimates. Another option is to ignore these bins altogether. For
readers that wish to do so, the numbers in each bin will be
given. However, we found that the quantity in bins with $>3$ objects
ranged from 0.51 to 1.41, with a mean 0.84 and standard deviation
0.24. On this basis, we chose to adopt $0.84\mu$ as our estimator for
$\sigma_0$ in bins with three or fewer objects.

\begin{figure*}
  \ForceWidth{10.0cm}
  \hSlide{-5cm}
  \BoxedEPSF{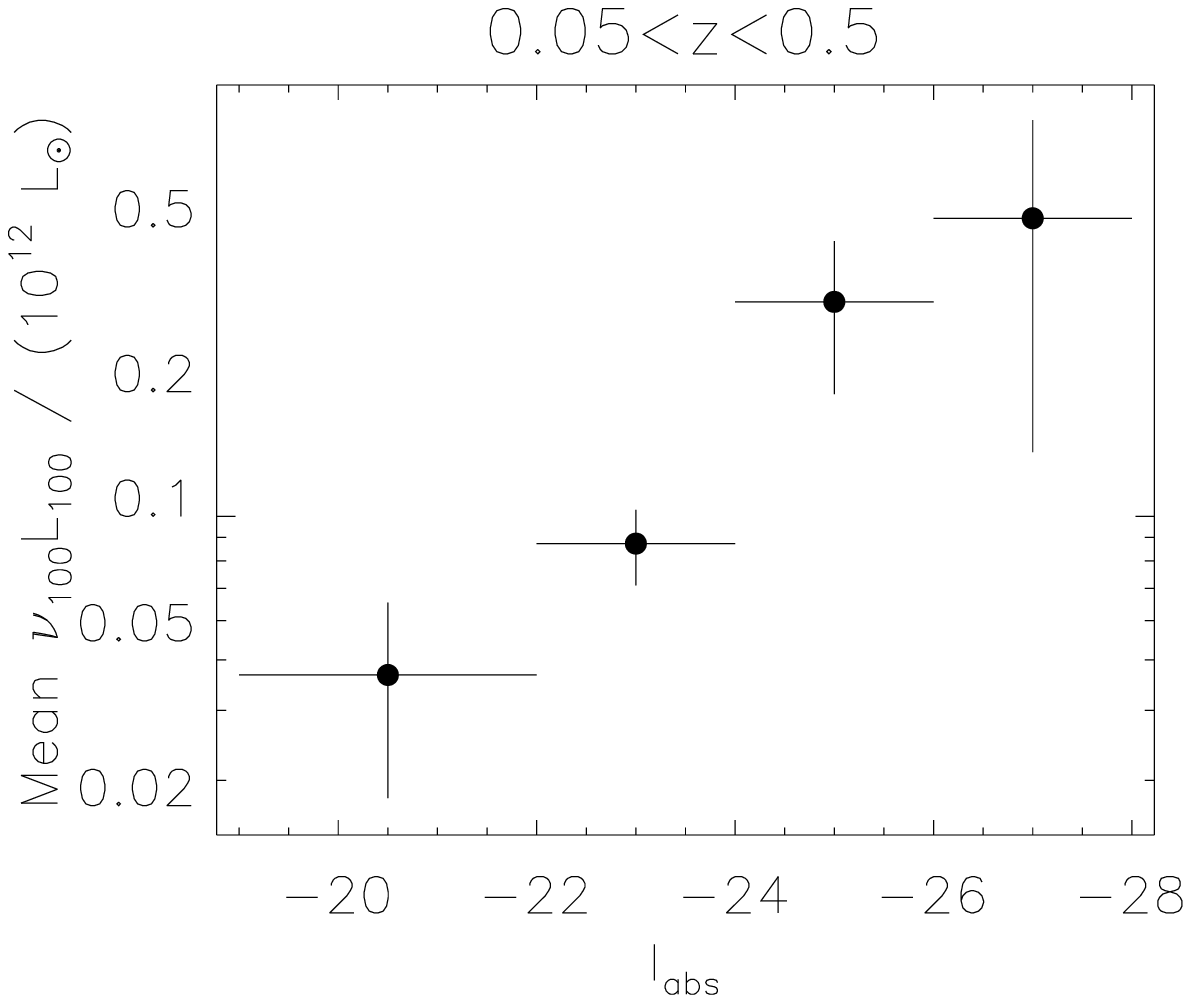}
  \vspace*{-7.2cm}
  \ForceWidth{10.0cm}
  \hSlide{5cm}
  \BoxedEPSF{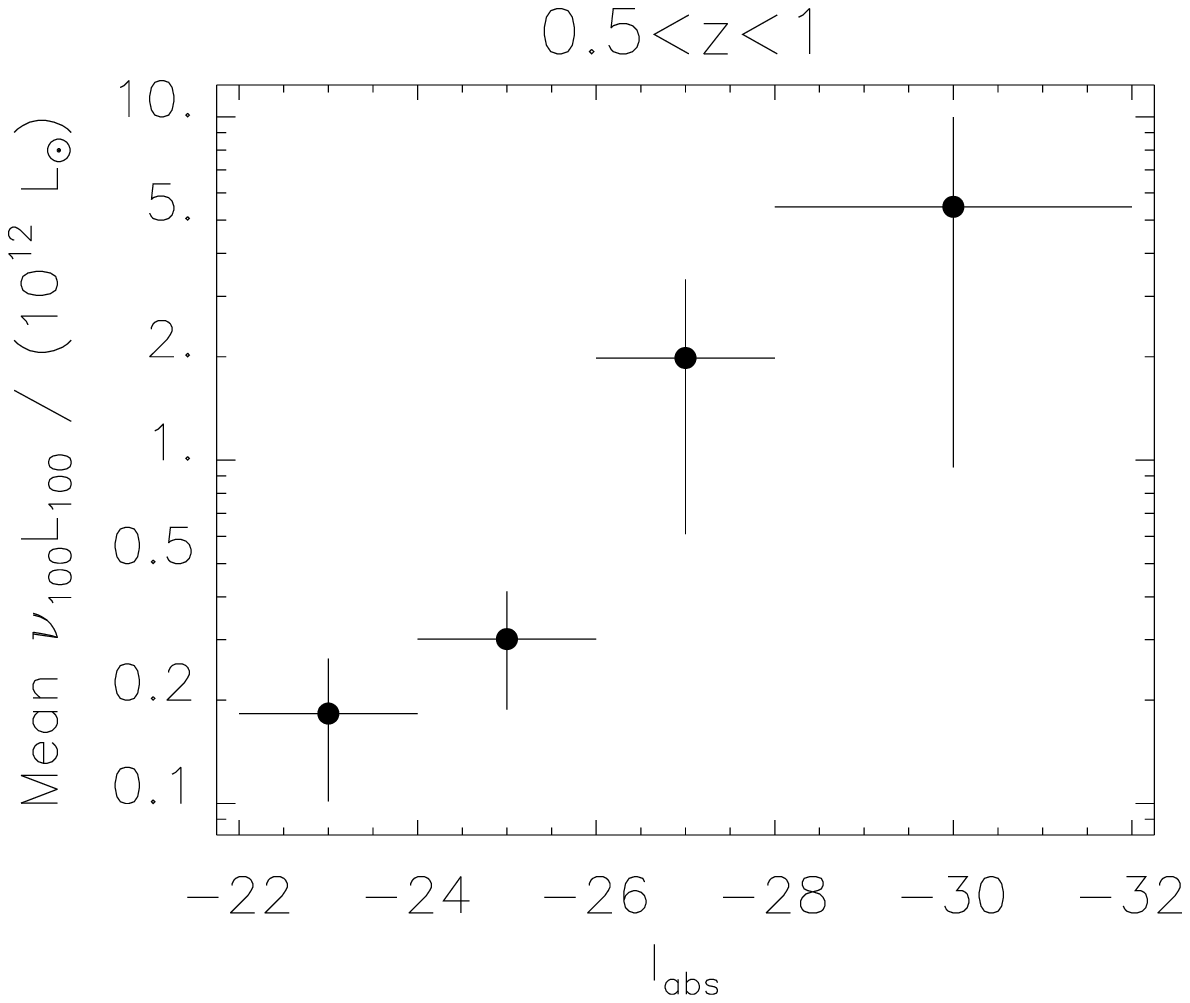}
\vspace*{-1.1cm}
%  \vspace*{-7.2cm}
  \ForceWidth{10.0cm}
  \hSlide{-5cm}
  \BoxedEPSF{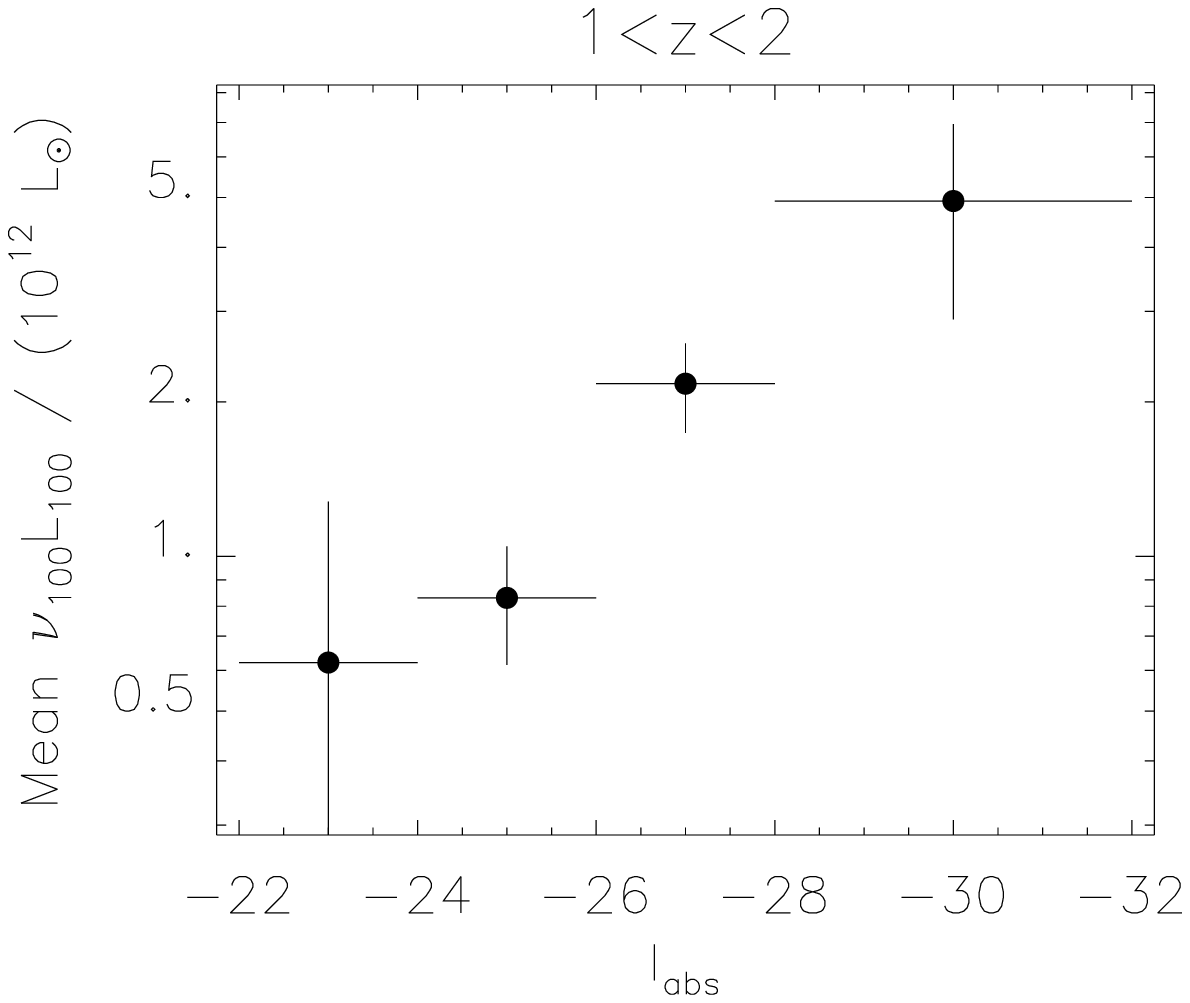}
  \vspace*{-7.2cm}
  \ForceWidth{10.0cm}
  \hSlide{5cm}
  \BoxedEPSF{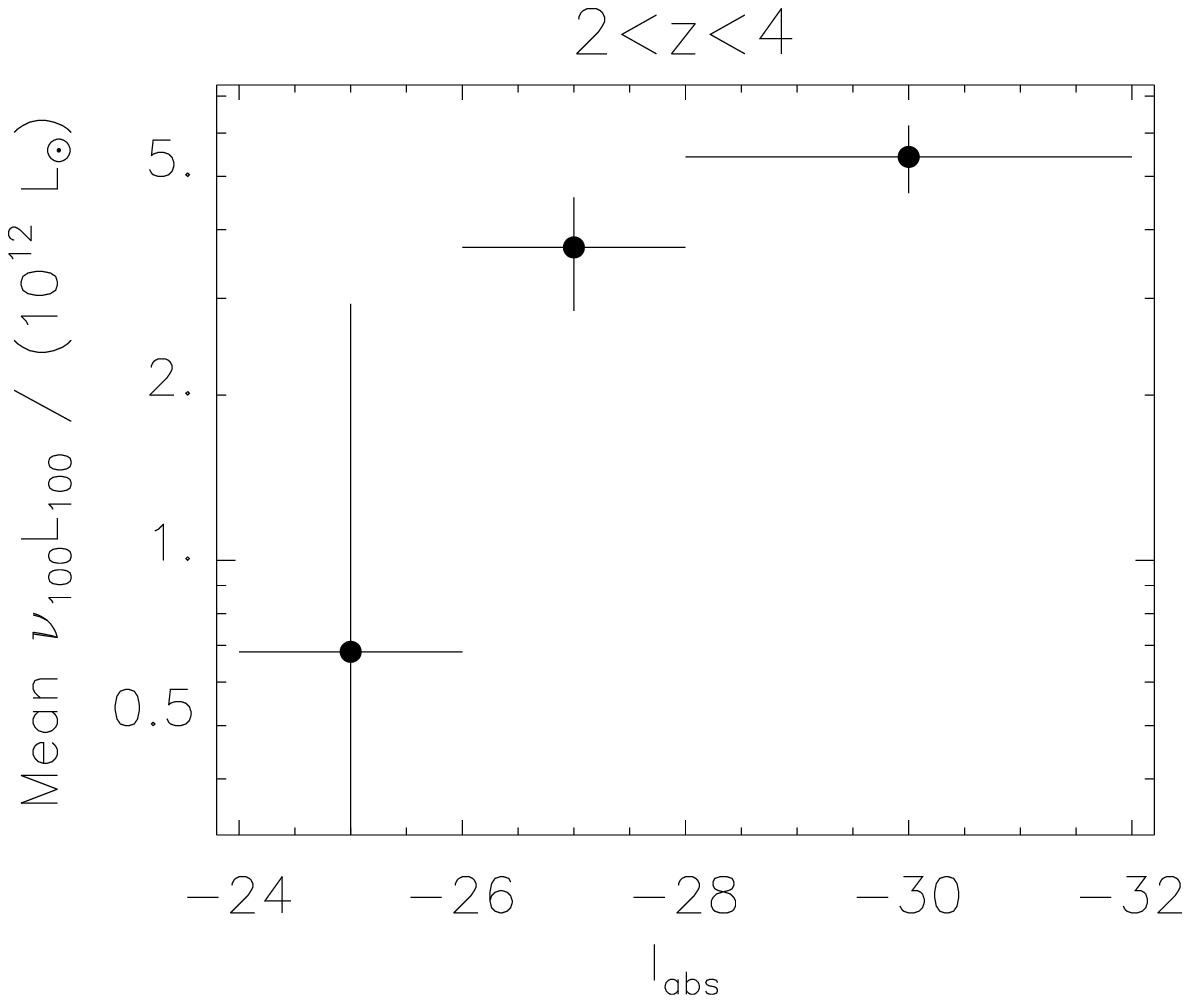}
\vspace*{-1.1cm}
  \ForceWidth{10.0cm}
  \hSlide{-5cm}
  \BoxedEPSF{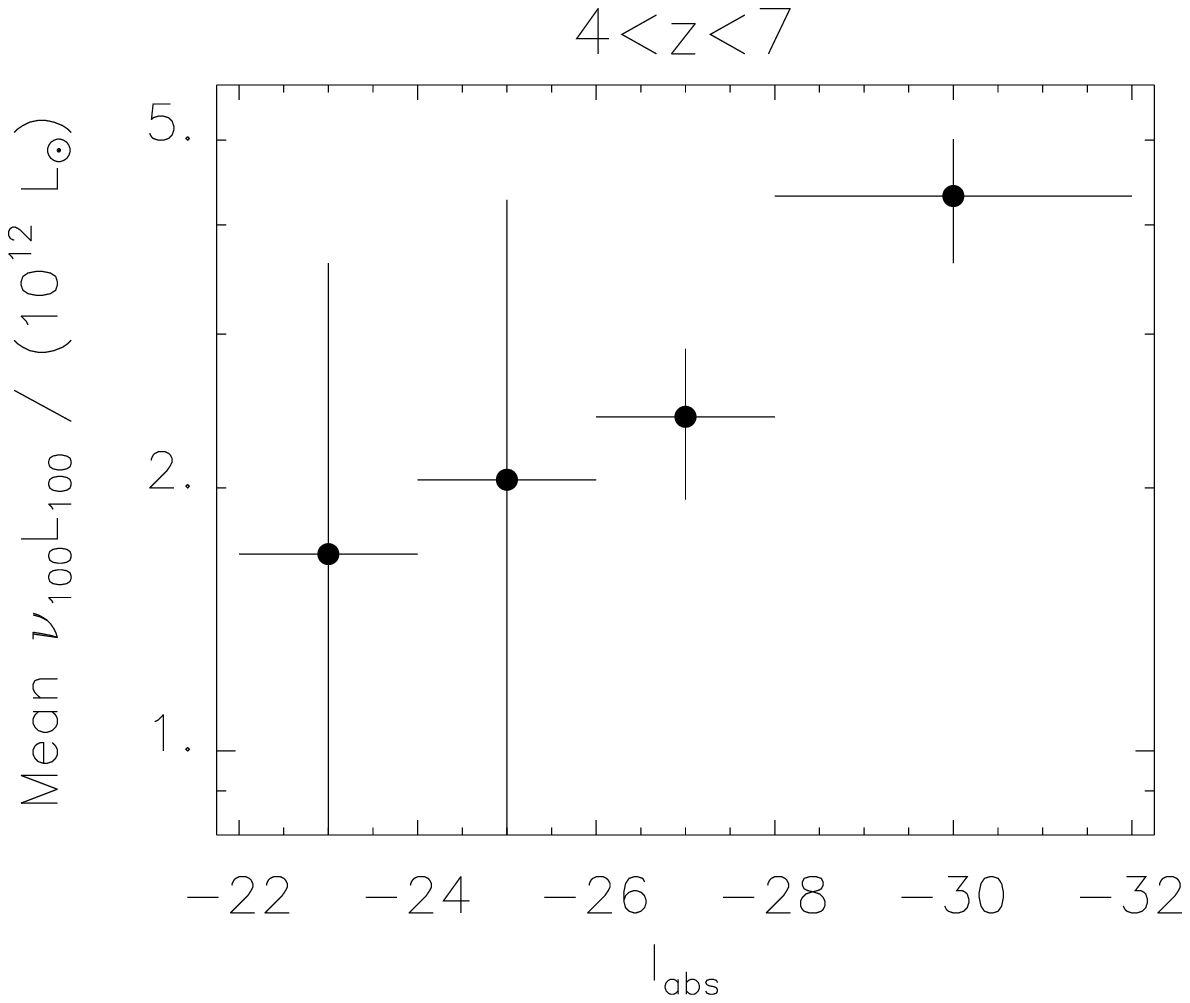}
\vspace*{-0.5cm}
  \caption{\label{fig:stacked_luminosity_vs_i}Optical luminosity
    dependence of the far-infrared luminosities of quasars, in five
    redshift slices, assuming an M82 SED template. Note that the
    far-infrared luminosities scale with the $I$-band
    luminosities, roughly as $L_{\rm FIR}\propto L_{\rm opt}^{0.5}$.}
\end{figure*}

\begin{figure*}
  \ForceWidth{10.0cm}
  \hSlide{-5cm}
  \BoxedEPSF{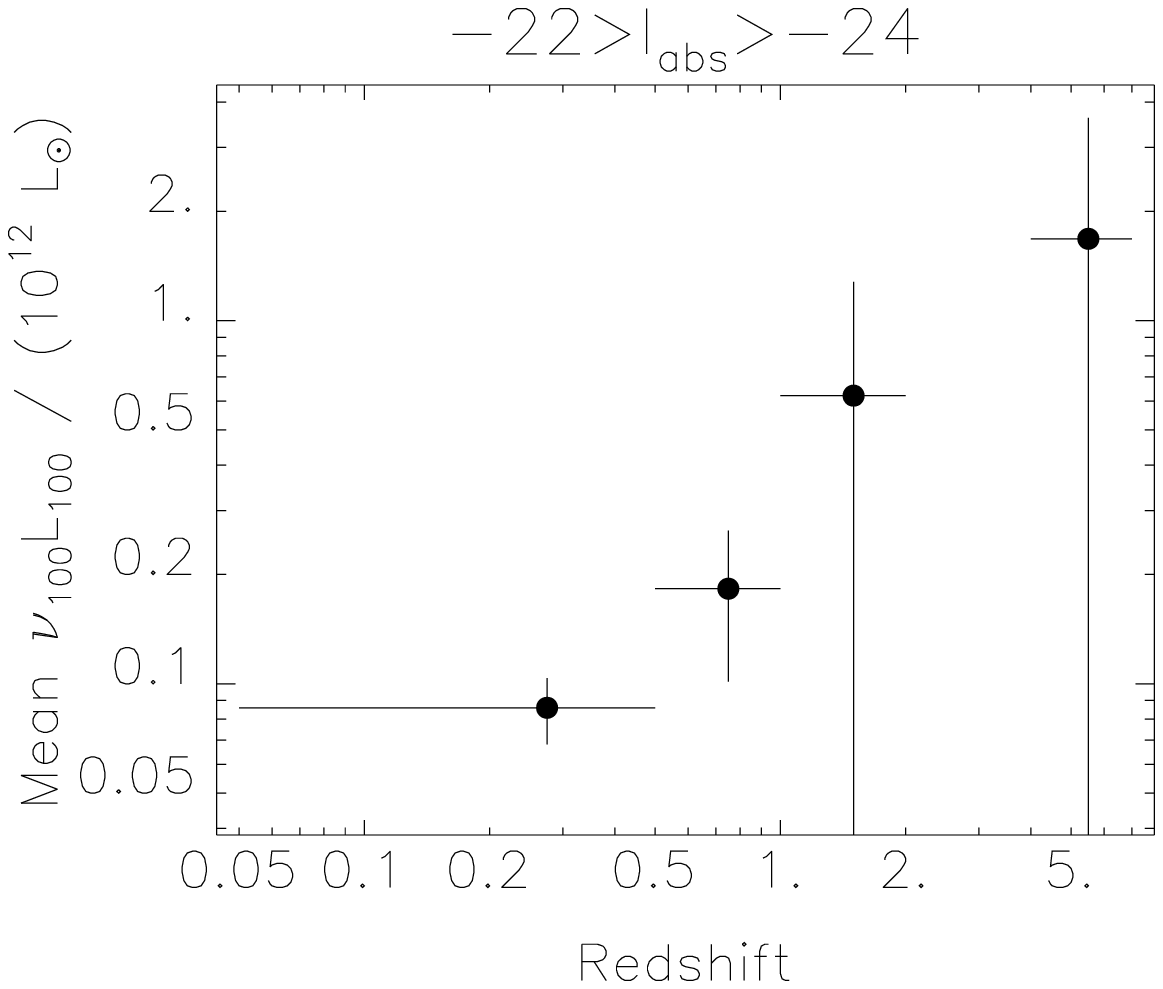}
  \vspace*{-7.2cm}
  \ForceWidth{10.0cm}
  \hSlide{5cm}
  \BoxedEPSF{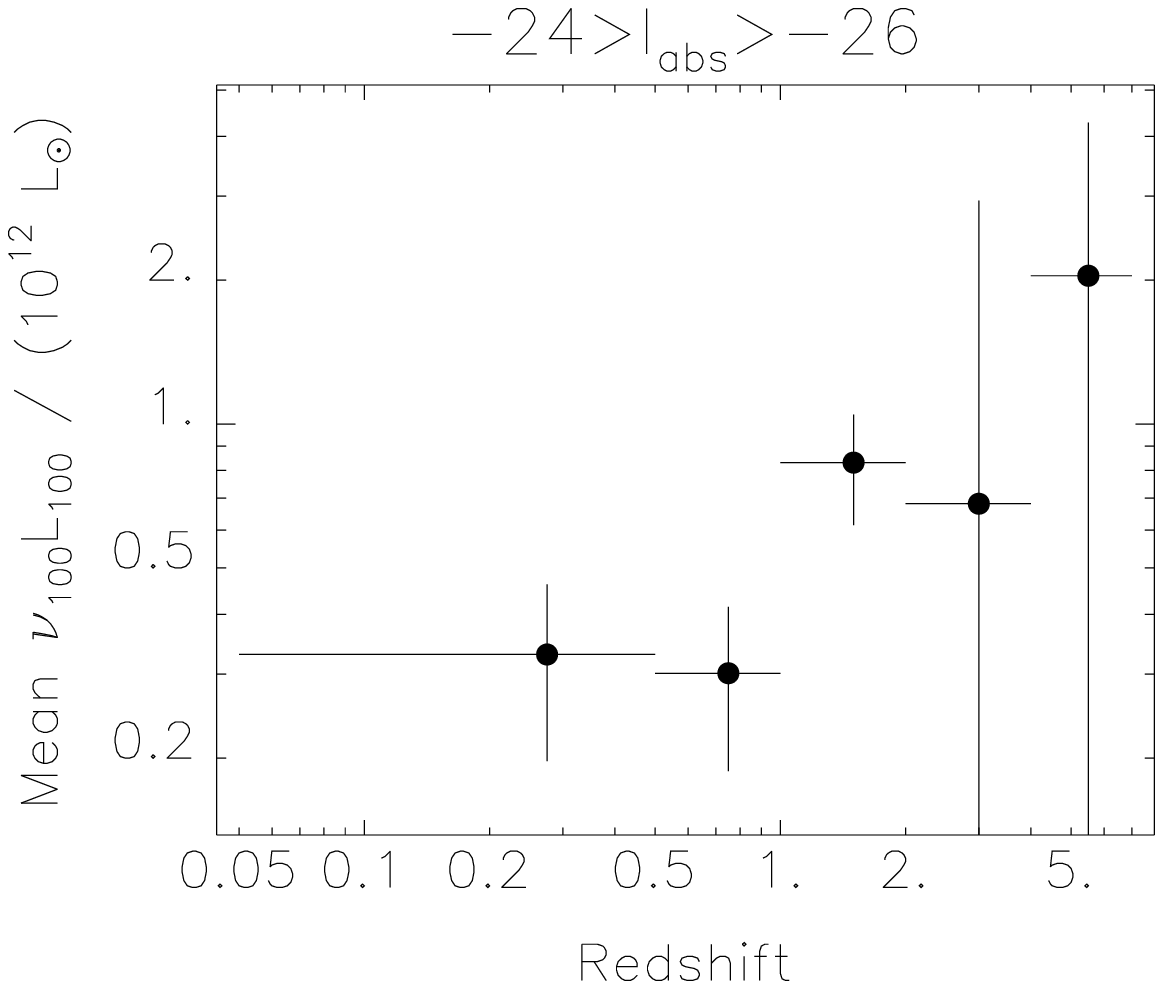}
\vspace*{-1.1cm}
%  \vspace*{-7.2cm}
  \ForceWidth{10.0cm}
  \hSlide{-5cm}
  \BoxedEPSF{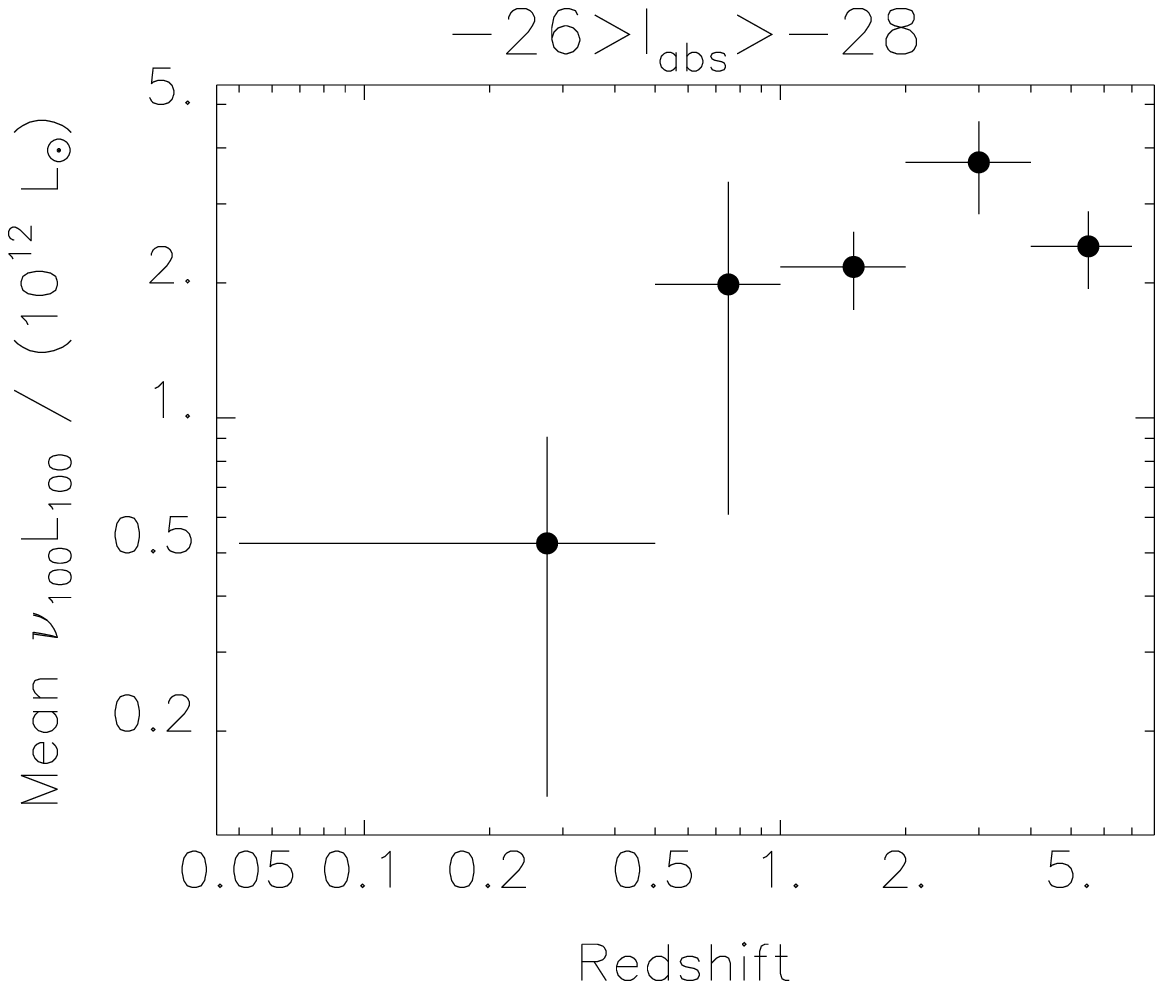}
  \vspace*{-7.2cm}
  \ForceWidth{10.0cm}
  \hSlide{5cm}
  \BoxedEPSF{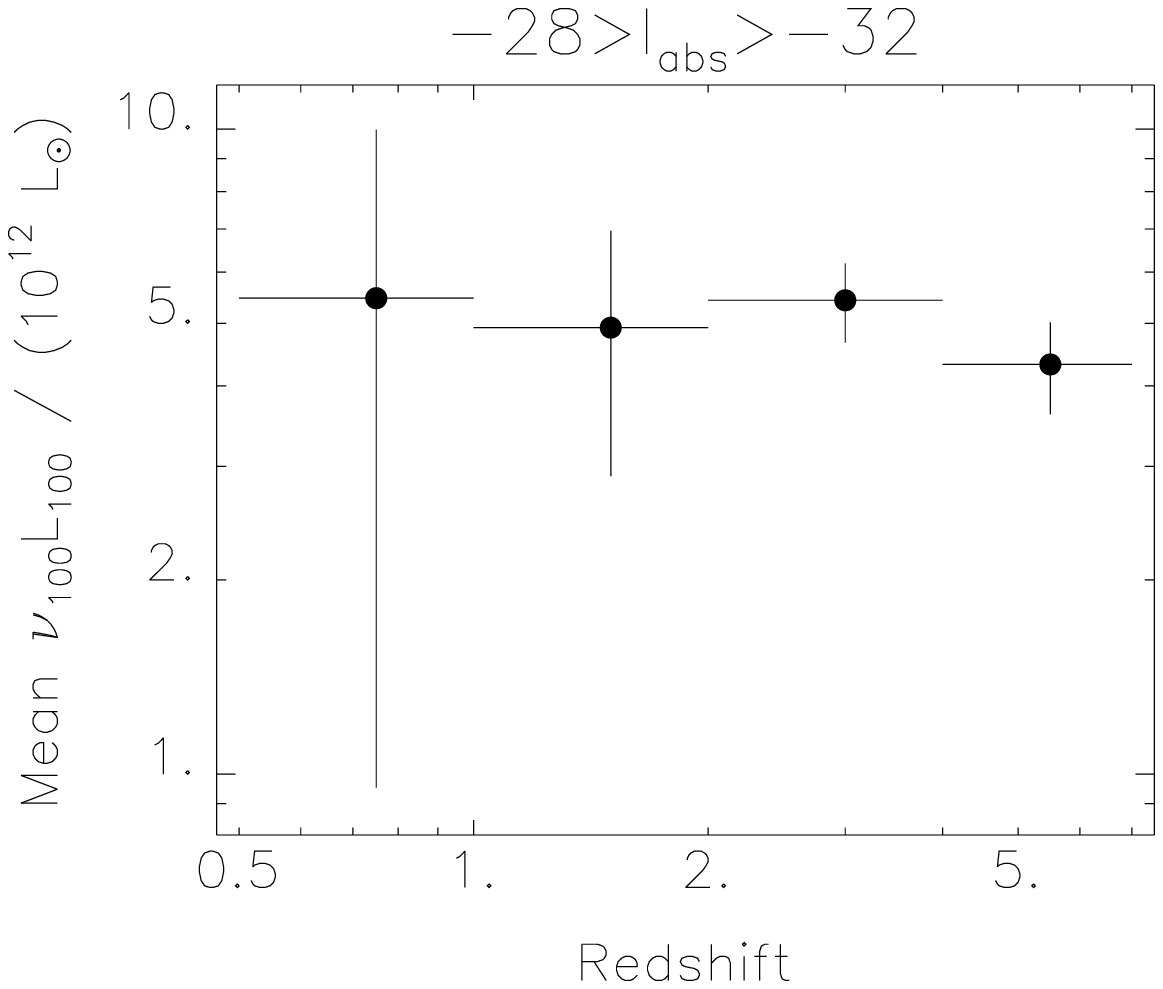}
\vspace*{-0.5cm}
  \caption{\label{fig:stacked_luminosity_vs_z}Evolution of quasar
    far-infrared luminosities in four absolute magnitude slices. Note
    the trends to brighter luminosities at higher redshifts, except in
    the brightest absolute magnitude strip. This can be interpreted as
    star formation in quasars decreasing with cosmic time, except
    perhaps in the most luminous quasars.}
\end{figure*}

We chose redshift and absolute magnitude bins and made a
noise-weighted stack of the starburst far-infrared luminosities of the
quasars in figure \ref{fig:iz} in each of these bins, using the
procedure above. Far-infrared luminosities were estimated assuming an
M82 SED. The results are listed in
table \ref{tab:m82_stacks}, and presented graphically in figures
\ref{fig:stacked_luminosity_vs_i} and
\ref{fig:stacked_luminosity_vs_z}. We will discuss the effects of
varying the SED in section 3.4. We used 160$\,\mu$m fluxes as our
far-infrared luminosity estimator for the SDSS quasars. For PG quasars
we used the 60$\,\mu$m fluxes at redshifts up to 0.3, and 100$\,\mu$m
fluxes at higher redshifts. Our results are not sensitive to the
choice of $z=0.3$ for this transition.

\begin{table*}
\begin{tabular}{llllll}
                        & $0.05<z<0.5$        & $0.5<z<1$           & $1<z<2$            & $2<z<4$            & $4<z<7$\\
$I_{\rm abs}<-28$	&                     & $5.5\pm4.5$ (2)     & $4.9\pm2.0$ (31)   & $5.43\pm0.76$ (66)   & $4.32\pm0.70$ (63)\\
$-26>I_{\rm abs}>-28$	& $0.52\pm0.38$ (9)   & $1.9\pm1.4$ (3)     & $2.17\pm0.42$ (48) & $3.71\pm0.87$ (76)   & $2.41\pm0.47$ (55)\\
$-24>I_{\rm abs}>-26$	& $0.33\pm0.13$ (28)  & $0.30\pm0.11$ (25)  & $0.83\pm0.22$ (74) & $0.7\pm2.3$ (3)    & $2.0\pm2.2$ (1)\\
$-22>I_{\rm abs}>-24$	& $0.086\pm0.018$ (55) & $0.18\pm0.08$ (20) & $0.62\pm0.66$ (2)  &                    & $1.7\pm1.9$ (1)\\
$I_{\rm abs}>-22$	& $0.041\pm0.021$ (3) &                     &                    &                    & \\
\end{tabular}
\caption{\label{tab:m82_stacks}
  Error-weighted mean far-infrared $\nu L_\nu$ luminosities of quasars 
  at $100\,\mu$m in redshift and
  luminosity bins, in units of $10^{12}L_\odot$ and assuming an M82 SED
  shape. Numbers in brackets are the number of quasars in each bin. The
  errors are $\left (\Sigma\sigma_i^{-2}\right )^{-1/2}$, as appropriate
  for error-weighted means, where the errors $\sigma_i$ include an
  estimate of the population dispersion.}
\end{table*}

\begin{table*}
\begin{tabular}{lllllll}
$z$ range & $p_1$ & $p_2$ & $\chi^2_\nu$ & Pr($\chi^2,\nu$) & $\chi^2_\nu$ & Pr($\chi^2,\nu$)\\
                    &       &       & & & \multicolumn{2}{c}{(no $I_{\rm abs}$ dependence)}\\
$0.05>z>0.5$&      28.3 $\pm$       1.5  &     -0.196 $\pm$     0.057 &     0.52  & 0.59 & 2.67 &     0.046\\
$0.5>z>1$&         27.00 $\pm$      0.81 &     -0.220 $\pm$     0.070 &     0.62  & 0.53 & 1.23 &      0.295\\
$1>z>2$&           25.21 $\pm$      0.53 &     -0.157 $\pm$     0.036 &     0.38  & 0.68 & 3.96 &    0.008\\
$2>z>4$&           19.4 $\pm$       5.5  &    -0.070 $\pm$     0.038 &     0.77  & 0.38 & 2.58 &     0.076\\
$z>4$&             22.1 $\pm$       2.9  &    -0.080 $\pm$     0.034 &     0.05  & 0.95 & 1.90 &      0.127\\
\end{tabular}
\caption{\label{tab:imag_fits}Fits to the data in figure 
\ref{fig:stacked_luminosity_vs_i}, assuming a functional form 
$\log_{10}(\nu_{100}L_{100}/10^{12}L_\odot)=p_2(I_{\rm abs}-p_1)$. 
The noise-weighted average value of $p_2$ at $z<2$ is $-0.176\pm0.028$, corresponding
to $\nu_{100}L_{100}\propto L_{\rm opt}^{0.441\pm0.069}$. 
Also given 
are the $\chi^2$ results for the best model with no $I_{\rm abs}$ dependence.}
\end{table*}

\begin{table*}
\begin{tabular}{lllllll}
$I_{\rm abs}$ range & $p_3$ & $p_4$ & $\chi^2_\nu$ & Pr($\chi^2,\nu$) & $\chi^2_\nu$ & Pr($\chi^2,\nu$)\\
                    &       &       & & & \multicolumn{2}{c}{(no evolution)}\\
$-22<I_{\rm abs}<-24$&    0.055 $\pm$     0.015 &       1.93 $\pm$       0.57 &  0.16 & 0.85 &    0.89 &      0.444\\
$-24<I_{\rm abs}<-26$&     0.180 $\pm$     0.077 &       1.43 $\pm$      0.54 &  0.60 & 0.61 &    1.39 &      0.234\\
$-26<I_{\rm abs}<-28$&     0.97 $\pm$      0.30 &      0.57 $\pm$        0.21 &  2.73 & 0.04 &    4.59 &    0.001\\
$-28<I_{\rm abs}<-32$&      7.3 $\pm$       3.5 &     -0.26 $\pm$        0.31 &  0.23 & 0.80 &    0.39 &      0.759\\
\end{tabular}
\caption{\label{tab:evol_fits}Fits to the data in figure 
\ref{fig:stacked_luminosity_vs_z}, assuming a functional form
$\nu_{100}L_{100}/10^{12}L_\odot=p_3(1+z)^{p_4}$. Also given 
are the $\chi^2$ results for the best non-evolving model.
The bin $-26<I_{\rm abs}<-28$ has a poor $\chi^2$ for the 
evolving model due to the highest-redshift bin, as discussed 
in section \ref{sec:stacked_luminosity_results}. Excluding
this data point, the average evolution at $I_{\rm abs}<-28$ is 
$(1+z)^{1.57\pm0.29}$. 
}
\end{table*}
  
\begin{table*}
\begin{tabular}{llllll}
                        & $0.05<z<0.5$        & $0.5<z<1$          & $1<z<2$            & $2<z<4$           & $4<z<7$\\
$10^5<M_{\rm BH}<10^6$    &                     &                    &                    & $2.4\pm3.6$ (1)   & $2.5\pm2.4$ (2)\\
$10^6<M_{\rm BH}<10^7$    & $0.017\pm0.012$ (2) &                    &                    & $9.3\pm8.6$ (1)     & $0.54\pm0.49$ (2)\\
$10^7<M_{\rm BH}<10^8$    & $0.25\pm0.13$ (21)  &                    &                    & $2.4\pm1.8$ (3)    & $2.8\pm1.1$ (6)\\
$10^8<M_{\rm BH}<10^9$    & $0.121\pm0.034$ (20)  &                  & $1.7\pm2.7$ (1)    & $5.4\pm1.0$ (11)    & $2.7\pm1.3$ (13)\\
$10^9<M_{\rm BH}<10^{10}$ & $0.37\pm0.35$ (3)   &  $10.0\pm9.0$ (1)  & $5.0\pm2.7$ (14)   & $2.8\pm1.1$ (15)    & $4.8\pm1.4$ (11)\\
\end{tabular}
\caption{\label{tab:m82_bh_stacks}
  Error-weighted mean far-infrared $\nu L_\nu$ luminosities of quasars 
  at $100\,\mu$m in redshift and
  black hole mass bins, in units of $10^{12}L_\odot$ and assuming an M82 SED
  shape. Numbers in brackets are the number of quasars in each bin. The
  errors are $\left (\Sigma\sigma_i^{-2}\right )^{-1/2}$, as appropriate
  for error-weighted means, where the errors $\sigma_i$ include an
  estimate of the population dispersion.}
\end{table*}

Taking bins in isolation, there are no high signal-to-noise
detections. Despite this however, there are trends apparent in
figures \ref{fig:stacked_luminosity_vs_i} and
\ref{fig:stacked_luminosity_vs_z}. Firstly, at all redshifts
there
appears to be a significant optical luminosity dependence, scaling
as $L_{\rm opt}^{0.441\pm0.069}$ at $z<2$ (see figure
\ref{fig:stacked_luminosity_vs_i} and table \ref{tab:imag_fits}) and
with a shallower scaling at higher redshifts. 
Secondly, at all absolute
magnitudes below $-28$ and redshifts $z<2$, there is a weaker signature of
evolution. 
(see figure \ref{fig:stacked_luminosity_vs_z} and 
table \ref{tab:evol_fits}). The poor $\chi^2$ for the power-law
evolution model in the $-26>I_{\rm abs}>-28$ bin is due to the highest
redshift data point, in which the positive evolution is reversed, 
curiously mirroring the evolution of quasar number density. If this
data point is excluded, the power-law fit parameters for this $I_{\rm abs}$ 
slice (table \ref{tab:evol_fits}) are 
$\nu_{100}L_{100}=(0.53\pm0.25)\times10^{12}L_\odot\times(1+z)^{1.46\pm0.43}$,
with $\chi^2=0.92$ and Pr$(\chi^2,\nu)=0.37$. This is consistent with the 
evolution at brighter absolute magnitudes, which justifies an estimate of 
the average evolution rate of $I_{\rm abs}>-28$ quasars of $(1+z)^{1.57\pm0.29}$. 

\subsection{Stacked black hole mass results}
\begin{figure*}
  \ForceWidth{10.0cm}
  \hSlide{-5cm}
  \BoxedEPSF{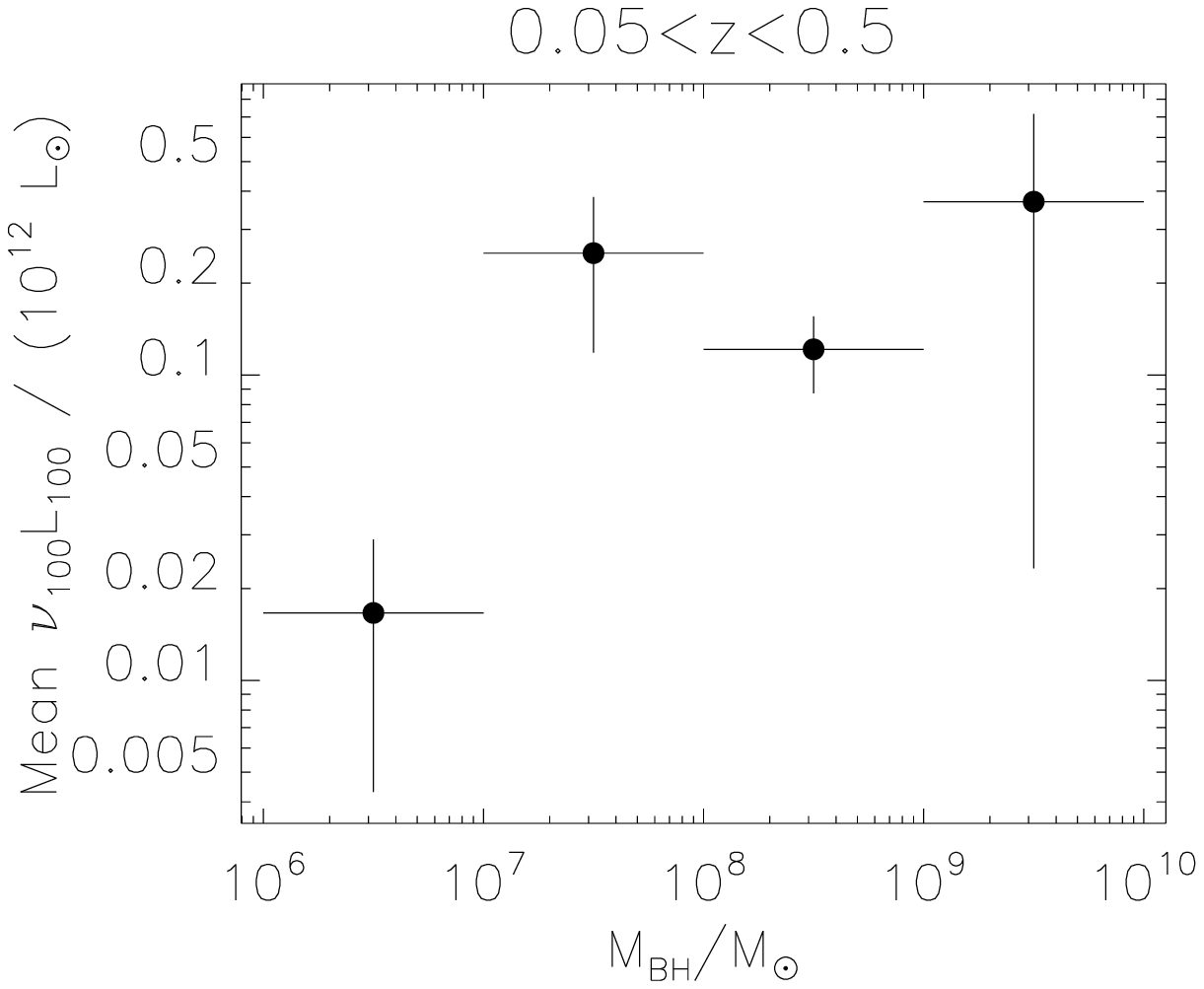}
  \vspace*{-7.2cm}
  \ForceWidth{10.0cm}
  \hSlide{5cm}
  \BoxedEPSF{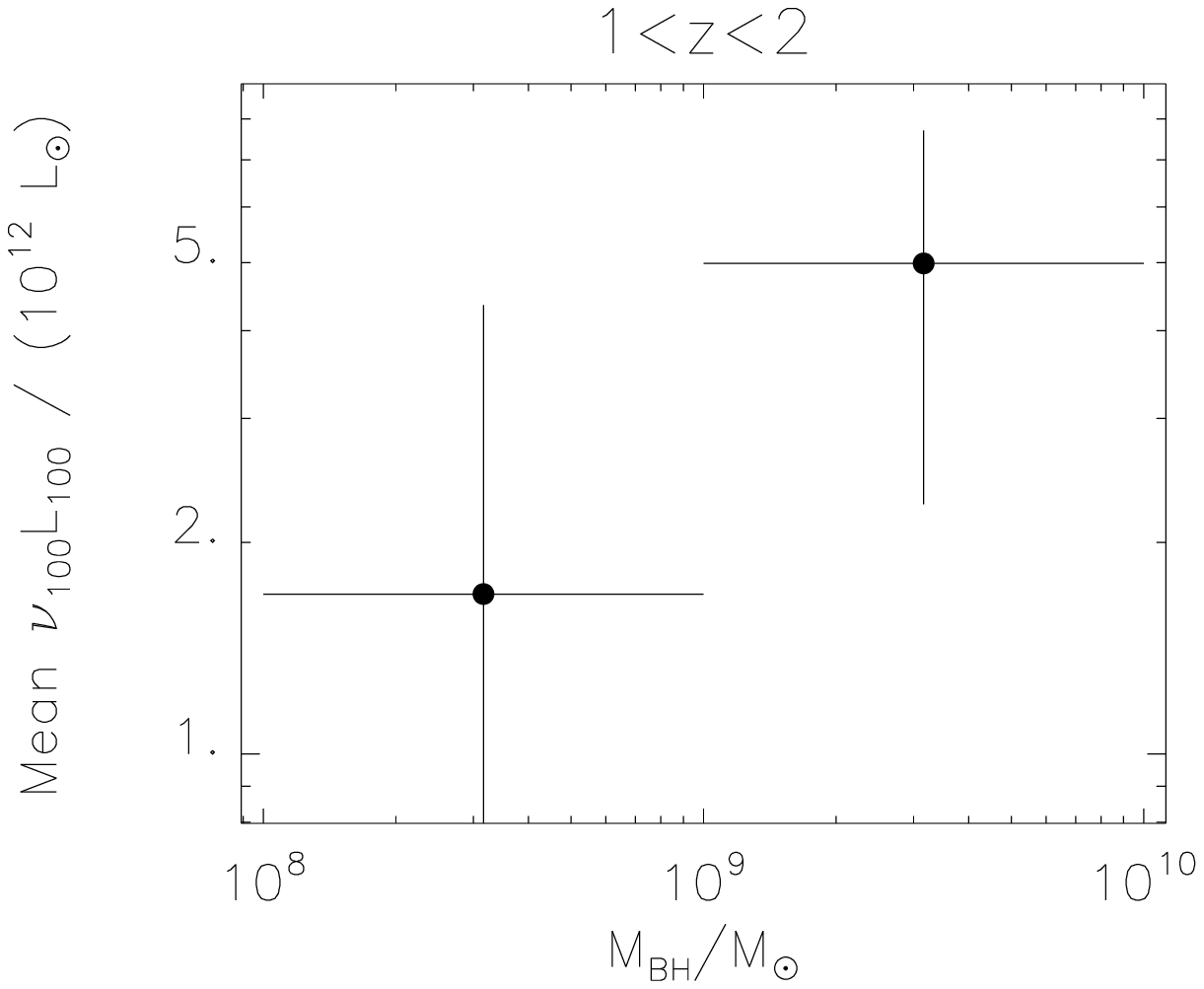}
\vspace*{-1.1cm}
%  \vspace*{-7.2cm}
  \ForceWidth{10.0cm}
  \hSlide{-5cm}
  \BoxedEPSF{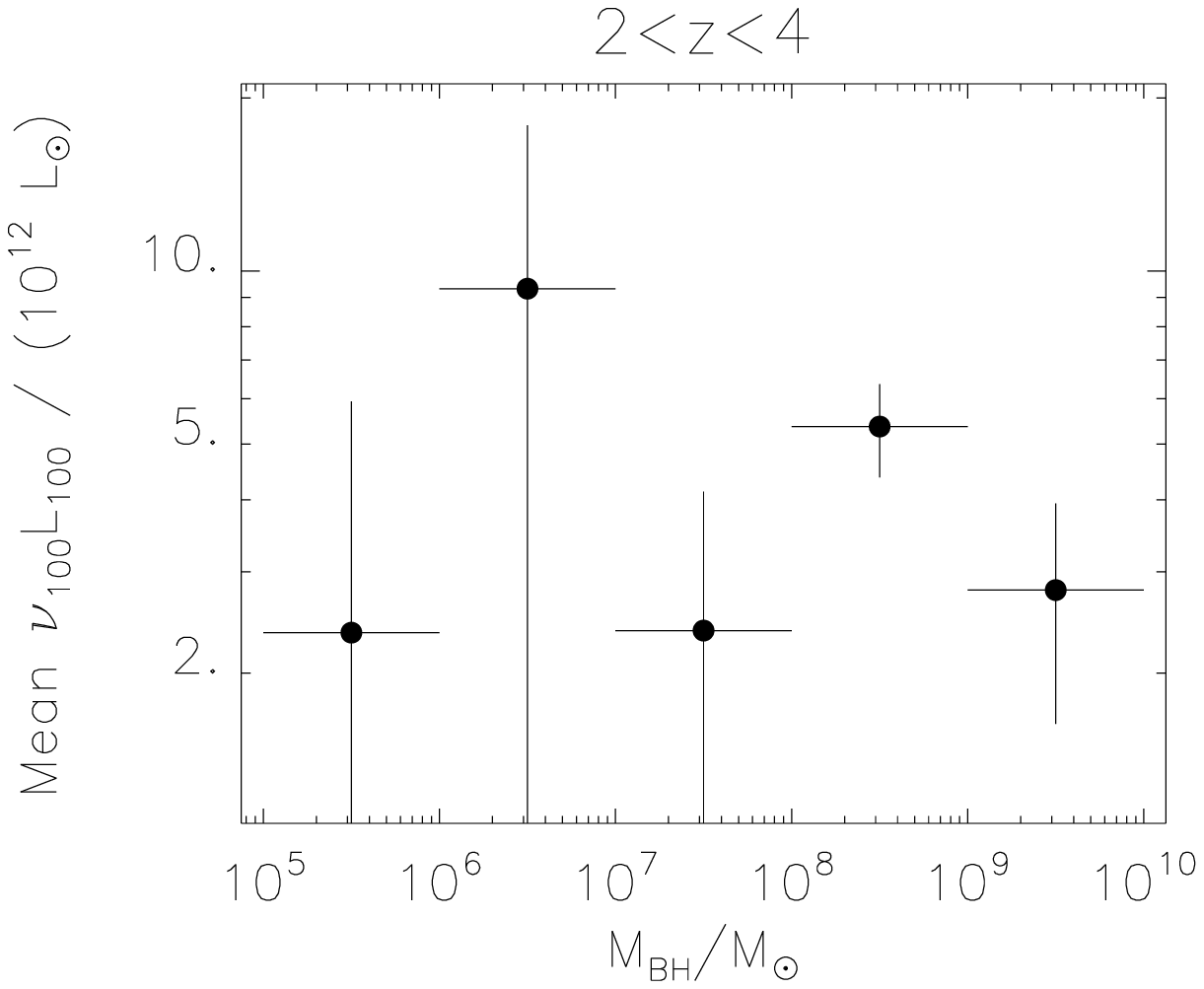}
  \vspace*{-7.2cm}
  \ForceWidth{10.0cm}
  \hSlide{5cm}
  \BoxedEPSF{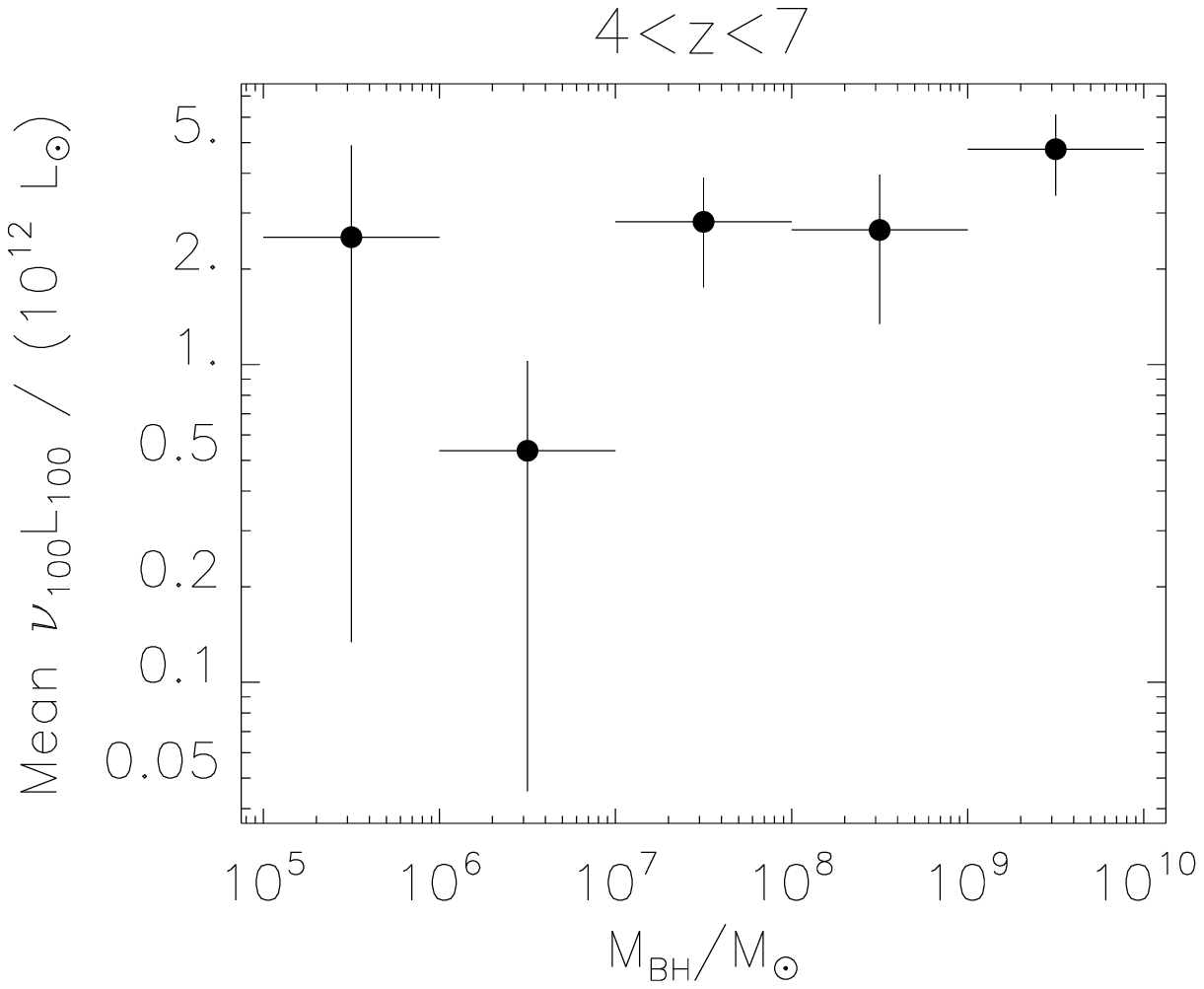}
\vspace*{-0.5cm}
\caption{\label{fig:stacked_luminosity_vs_bh}Host galaxy
  far-infrared luminosities as a function of black hole mass, for
  selected bins in table \ref{tab:m82_bh_stacks}}
\end{figure*}

\begin{figure*}
  \ForceWidth{10.0cm}
  \hSlide{-5cm}
  \BoxedEPSF{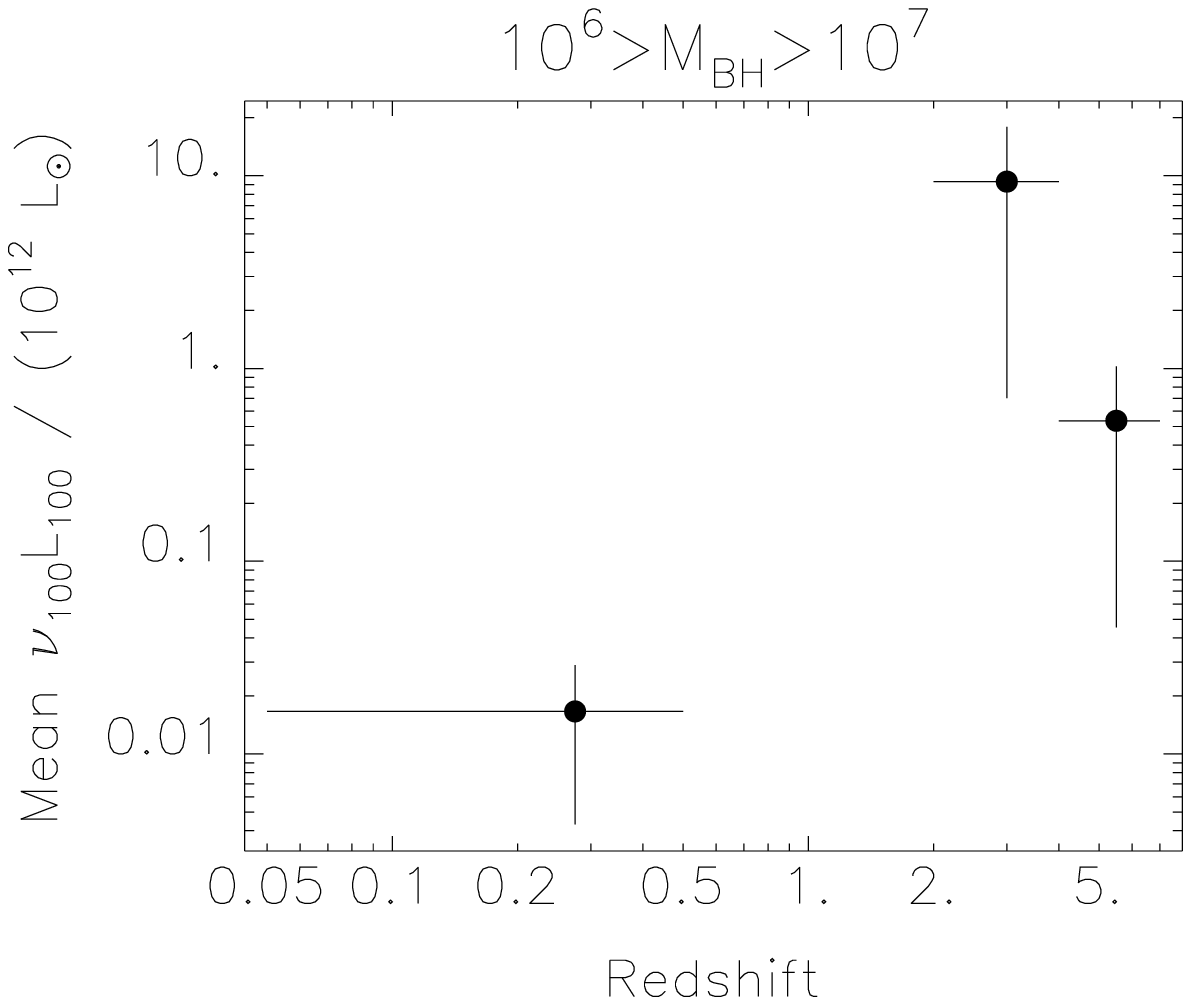}
  \vspace*{-7.2cm}
  \ForceWidth{10.0cm}
  \hSlide{5cm}
  \BoxedEPSF{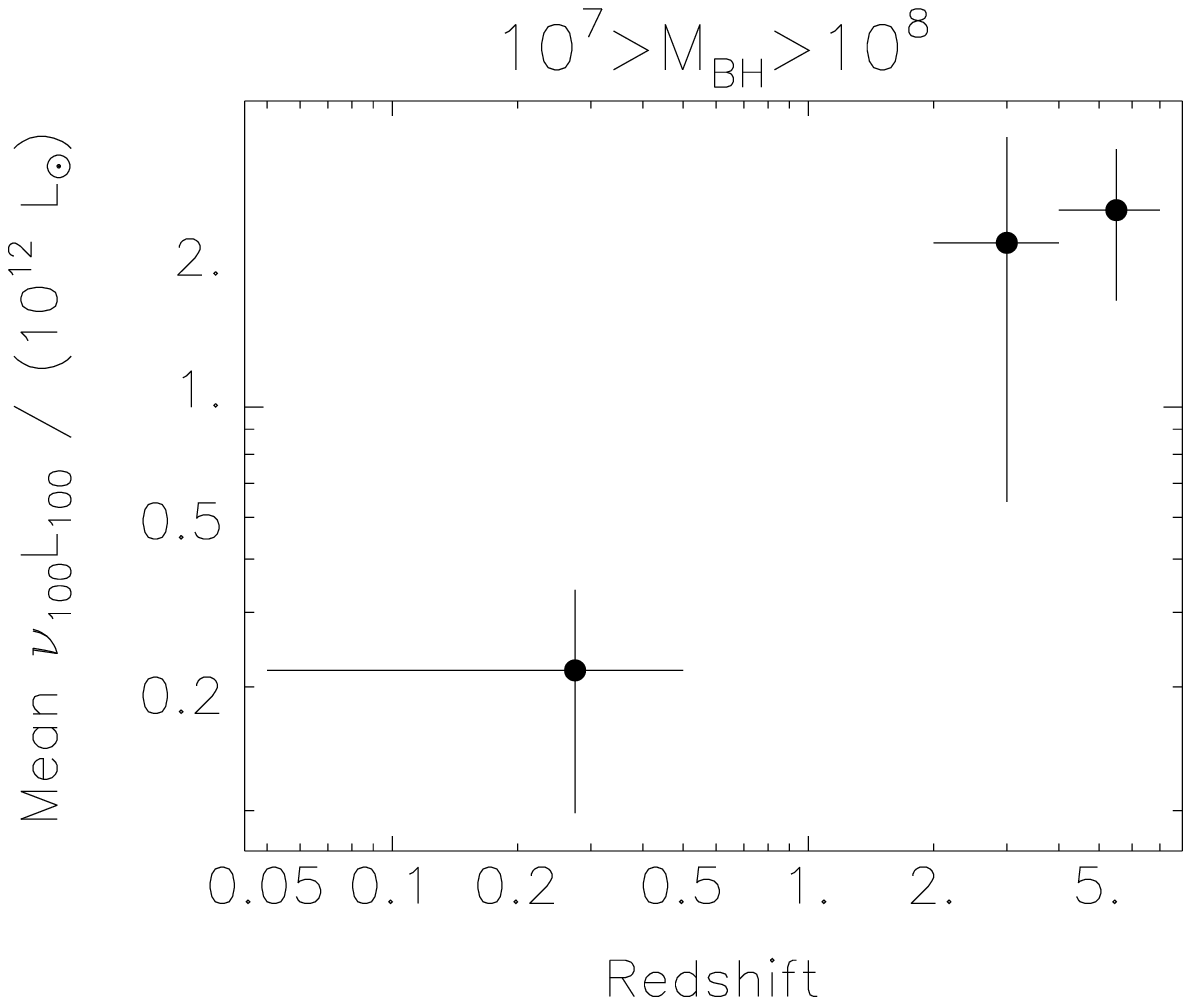}
\vspace*{-1.1cm}
%  \vspace*{-7.2cm}
  \ForceWidth{10.0cm}
  \hSlide{-5cm}
  \BoxedEPSF{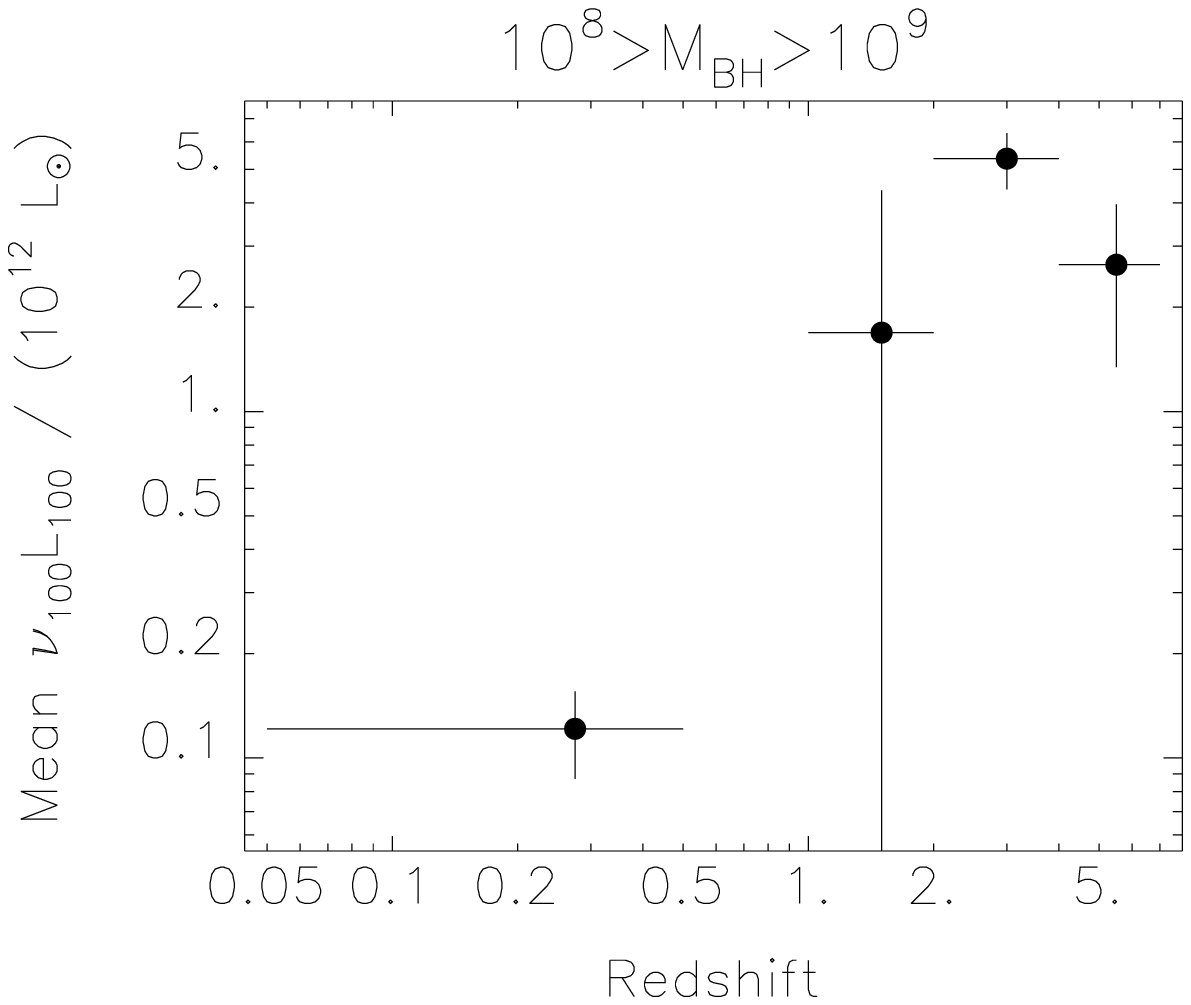}
  \vspace*{-7.2cm}
  \ForceWidth{10.0cm}
  \hSlide{5cm}
  \BoxedEPSF{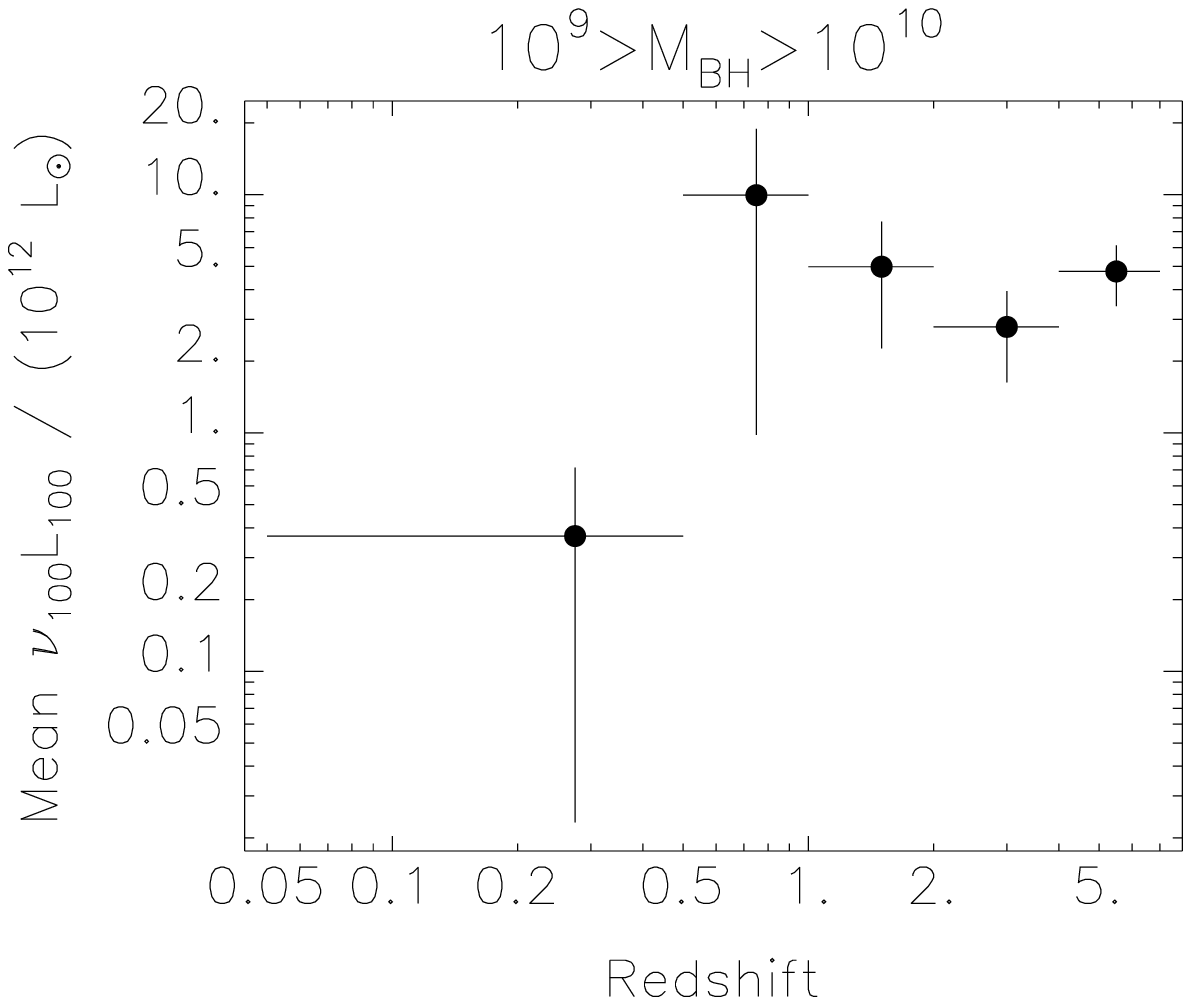}
\vspace*{-0.5cm}
\caption{\label{fig:stacked_luminosity_vs_z_bh}Host galaxy
  far-infrared luminosities as a function of redshift, for
  selected bins in table \ref{tab:m82_bh_stacks}}
\end{figure*}

We next examined our data for trends with black hole mass. The black
hole mass computation is based on the extrapolation of the
reverberation mapping technique, which considers the velocity
(full-width half maximum) of emission lines and relates the size of
the Broad Line Region (BLR) to the continuum luminosity. 
Assuming the dynamics of the BLR are dominated by the gravity
of the black hole, the black hole mass is then expressed as
\begin{equation}
M_{\rm BH}\simeq R_{\rm BLR}\times v^2/G
\label{eqn:mbh}
\end{equation}
where $R_{\rm BLR}$ is the radius of the BLR and $v$ is the velocity
of the emission line gas. The velocity $v$ is estimated from the
FWHM of H$_\beta$, MgII or CIV depending on the redshift 
(Kapsi et al. 2000). For
quasars with redshifts greater than $z\simeq0.8$, H$_\beta$
is not present in optical spectra, and MgII and
CIV have been suggested as alternative estimators. For a detailed
analysis of the method and the use of the various emission lines see
Kapsi et al. 2000, McLure \& Dunlop 2004, and Warner et al. 2004.
Depending on the redshift of the object and up to a redshift of 4.8,
the following relations are used, that have been argued by these
authors to provide equivalent estimates:
\begin{equation}
0.0<z<0.8: \frac{M_{\rm BH}}{M_\odot}=4.7\left (\frac{L_{5100}}{10^{37}\rm
  W}\right )^{0.61}\left (\frac{{\rm FWHM}({\rm H}_\beta)}{\rm
  km/s}\right )^2 
\label{eqn:mbh2}
\end{equation}
\begin{equation}
0.8<z<2.1: \frac{M_{\rm BH}}{M_\odot}=3.2\left (\frac{L_{3000}}{10^{37}\rm
  W}\right )^{0.62}\left (\frac{{\rm FWHM}({\rm MgII})}{\rm
  km/s}\right )^2 
\label{eqn:mbh3}
\end{equation}
\begin{equation}
2.1<z<4.8: \frac{M_{\rm BH}}{M_\odot}=1.4\left (\frac{L_{1450}}{10^{37}\rm
  W}\right )^{0.70}\left (\frac{{\rm FWHM}({\rm CIV})}{\rm km/s}\right
)^2 
\label{eqn:mbh4}
\end{equation}
where $L_\lambda$ is the luminosity at wavelength $\lambda$. Black
hole masses were estimated for objects with available SDSS DR5
spectroscopy. The FWHM of the lines were
derived from the emission line sigmas given by the SDSS pipeline,
present in the second extension of the fits spectra. The error bars on
the black hole masses are estimated from the uncertainty of the FWHM
of the lines. We also searched the literature for additional black
hole mass estimates. For Palomar-Green quasars in Boroson \& Green
(1992), we recomputed the black hole masses using the relations
above. For the $z=6.28$ quasar SDSS\,J1148+5251, we applied equation
\ref{eqn:mbh2} to the data in Shields et al. (2006). For other $z>5$
quasars, black hole masses were taken from Kurk et al. (2007) and
Jiang et al. (2007), using our equation \ref{eqn:mbh3} where possible,
and equation \ref{eqn:mbh4} where only CIV is available. The black
hole mass estimates vary between these two groups, even when using
consistent conversions based on the same emission line, but because
our mass bins are very broad (1 dex), these fractional variations
($\sim50\%$) are not enough to move quasars to different bins.  The
results of the stacks in black hole mass and redshift bins are given
in table \ref{tab:m82_bh_stacks}, and plotted in figures
\ref{fig:stacked_luminosity_vs_bh} and
\ref{fig:stacked_luminosity_vs_z_bh}. These figures give the
host galaxy far-infrared luminosities as a function of black hole
mass and of redshift. Tables \ref{tab:mbh_fits} and 
\ref{tab:evol_fits_bh} demonstrate the statistical significances of
the trends in these figures. 

\begin{table*}
\begin{tabular}{lllllll}
$z$ range & $p_5$ & $p_6$ & $\chi^2_\nu$ & Pr($\chi^2,\nu$) & $\chi^2_\nu$ & Pr($\chi^2,\nu$)\\
                    &       &       & & & \multicolumn{2}{c}{(no $I_{\rm abs}$ dependence)}\\
$0.05>z>0.5$&     0.42 $\pm$      0.14  &      -10.66 $\pm$      0.86              &       1.20 &      0.302  &       4.01 &    0.007\\
$2>z>4$     &  0.000 $\pm$     0.082    &       $3\times10^3$ $\pm$  $1\times10^6$  &       1.48 &      0.217 &       1.11 &      0.348\\
$z>4$       &     0.247 $\pm$     0.082 &      -6.70 $\pm$      0.75                &       0.87 &      0.454 &       3.03 &     0.017\\
\end{tabular}
\caption{\label{tab:mbh_fits}Fits to the data in figure 
\ref{fig:stacked_luminosity_vs_bh}, assuming a functional form 
$\log_{10}(\nu_{100}L_{100}/10^{12}L_\odot)=p_5(\log_{10}(M_{\rm BH})+p_6)$. Also given 
are the $\chi^2$ results for the best model with no $M_{\rm BH}$ dependence.}
\end{table*}
  
\begin{table*}
\begin{tabular}{lllllll}
$M_{\rm BH}$ range & $p_7$ & $p_8$ & $\chi^2_\nu$ & Pr($\chi^2,\nu$) & $\chi^2_\nu$ & Pr($\chi^2,\nu$)\\
                    &       &       & & & \multicolumn{2}{c}{(no evolution)}\\
$10^{6}<M_{\rm BH}<10^{7}$&   0.0099 $\pm$    0.0085 &       2.14 $\pm$      0.71 &       1.12 &     0.289 &       1.14 &      0.319\\
$10^{7}<M_{\rm BH}<10^{8}$&     0.18 $\pm$      0.11 &       1.49 $\pm$      0.38 &      0.31 &      0.581 &       3.55 &     0.029\\
$10^{8}<M_{\rm BH}<10^{9}$&    0.082 $\pm$     0.024 &       2.12 $\pm$      0.24 &       8.47 &     0.0002 &       10.71 &   $4\times10^{-7}$\\
$10^{9}<M_{\rm BH}<10^{10}$&     0.36 $\pm$     0.26 &       1.40 $\pm$      0.43 &       1.04 &      0.374 &       4.09 &    0.003\\
\end{tabular}
\caption{\label{tab:evol_fits_bh}Fits to the data in figure 
\ref{fig:stacked_luminosity_vs_z_bh}, assuming a functional form
$\nu_{100}L_{100}/10^{12}L_\odot=p_7(1+z)^{p_8}$. Also given 
are the $\chi^2$ results for the best non-evolving model.}
\end{table*}

\subsection{Robustness to SED assumptions}
Any multi-wavelength compilation such as this will inevitably rely on
SED assumptions to relate the multi-wavelength observations. We have
assumed an M82 SED up to this point, and quoted far-infrared
luminosities on that basis, but this will inevitably neglect any AGN
dust tori contributions in the mid-infrared. At best, our far-infrared
luminosities can only be considered estimates of the starburst
bolometric contributions. Starbursts, too, have a variety of SEDs, and
in this section we will test the robustness to our assumed starburst
SED shape. The most far-infrared-luminous quasars (i.e. those with
direct SWIRE detections) were found by
Hatziminaoglou et al. (2008) to resemble the heavily-obscured
starburst Arp 220 more often than M82, though they conjectured that
fainter quasars would be more likely to resemble M82. We tested our
SED dependence by re-running our stacking analyses with an Arp 220
spectrum. Reassuringly, very
similar trends are present as in the M82 case, perhaps because most of
the rest-frame measurements are in spectral regions where the M82 and
Arp 220 SEDs are similar (see tables \ref{tab:arp220_stacks} and
\ref{tab:arp220_bh_stacks}). Moreover, it would not appear to be possible
to attribute the trends in figures \ref{fig:stacked_luminosity_vs_i},
\ref{fig:stacked_luminosity_vs_z}, \ref{fig:stacked_luminosity_vs_bh}
and \ref{fig:stacked_luminosity_vs_z_bh} to variations in SED shape as
a function of quasar absolute magnitude, black hole mass or redshift.

\begin{table*}
\begin{tabular}{llllll}
                        & $0.05<z<0.5$        & $0.5<z<1$           & $1<z<2$            & $2<z<4$            & $4<z<7$\\
$I_{\rm abs}<-28$	&                     & $8.4\pm6.9$ (2)     & $6.2\pm2.8$ (31)   & $3.95\pm0.52$ (66)   & $3.96\pm0.64$ (63)\\
$-26>I_{\rm abs}>-28$	& $0.53\pm0.38$ (9)   & $2.3\pm1.6$ (3)     & $2.51\pm0.48$ (48) & $4.4\pm1.1$ (76)   & $2.19\pm0.43$ (55)\\
$-24>I_{\rm abs}>-26$	& $0.47\pm0.20$ (28)  & $0.29\pm0.11$ (25)  & $0.84\pm0.21$ (74) & $1.0\pm3.5$ (3)    & $2.0\pm2.2$ (1)\\
$-22>I_{\rm abs}>-24$	& $0.107\pm0.021$ (55) & $0.18\pm0.08$ (20) & $0.59\pm0.63$ (2)  &                    & $1.7\pm2.0$ (1)\\
$I_{\rm abs}>-22$	& $0.055\pm0.028$ (3) &                     &                    &                    & \\
\end{tabular}
\caption{\label{tab:arp220_stacks} 
  Error-weighted mean far-infrared $\nu L_\nu$ luminosities of quasars 
  at $100\,\mu$m in redshift and
  luminosity bins, in units of $10^{12}L_\odot$ and assuming an Arp\,220 SED
  shape. Numbers in brackets are the number of quasars in each bin. The
  errors are $\left (\Sigma\sigma_i^{-2}\right )^{-1/2}$, as appropriate
  for error-weighted means, where the errors $\sigma_i$ include an
  estimate of the population dispersion.}
\end{table*}

\begin{table*}
\begin{tabular}{llllll}
                        & $0.05<z<0.5$        & $0.5<z<1$          & $1<z<2$            & $2<z<4$           & $4<z<7$\\
$10^5<M_{\rm BH}<10^6$    &                     &                    &                    & $1.8\pm2.7$ (1)   & $2.4\pm2.3$ (2)\\
$10^6<M_{\rm BH}<10^7$    & $0.023\pm0.017$ (2) &                    &                    & $7.2\pm6.6$ (1)     & $0.46\pm0.42$ (2)\\
$10^7<M_{\rm BH}<10^8$    & $0.35\pm0.19$ (21)  &                    &                    & $1.6\pm1.1$ (3)    & $2.38\pm0.90$ (6)\\
$10^8<M_{\rm BH}<10^9$    & $0.183\pm0.053$ (20)  &                  & $1.0\pm1.6$ (1)    & $4.25\pm0.80$ (11)    & $2.5\pm1.2$ (13)\\
$10^9<M_{\rm BH}<10^{10}$ & $0.37\pm0.34$ (3)   &  $15\pm14$ (1)  & $6.0\pm3.8$ (14)   & $2.16\pm0.86$ (15)    & $4.6\pm1.3$ (11)\\
\end{tabular}
\caption{\label{tab:arp220_bh_stacks}
  Error-weighted mean far-infrared $\nu L_\nu$ luminosities of quasars 
  at $100\,\mu$m in redshift and
  black hole mass bins, in units of $10^{12}L_\odot$ and assuming an Arp\,220 SED
  shape. Numbers in brackets are the number of quasars in each bin. The
  errors are $\left (\Sigma\sigma_i^{-2}\right )^{-1/2}$, as appropriate
  for error-weighted means, where the errors $\sigma_i$ include an
  estimate of the population dispersion.}
\end{table*}

Few objects in our compilation have photometry at more than one
wavelength, except the SWIRE quasars. We tested our SED assumption by
fitting M82 and Arp 220 SEDs to the SWIRE photometry. Of 236 quasars
with photometric measurements at both 70$\,\mu$m and 160$\,\mu$m, 167
had $\chi^2<1$ for either M82 or Arp 220 SEDs. Of these, 96/167 (57\%)
had a lower $\chi^2$ for M82. Since most of this photometry is
non-detections, all this is capable of showing is that the underlying
mean SED is more likely to resemble M82 than Arp 220, in agreement
with the suggestion of Hatziminaoglou et al. (2008). Objects with
$\chi^2>1$ typically had evidence of a mid-infrared excess, suggestive
of an AGN dust torus.

\begin{figure*}
  \ForceWidth{10.0cm}
  \hSlide{-4cm}
  \BoxedEPSF{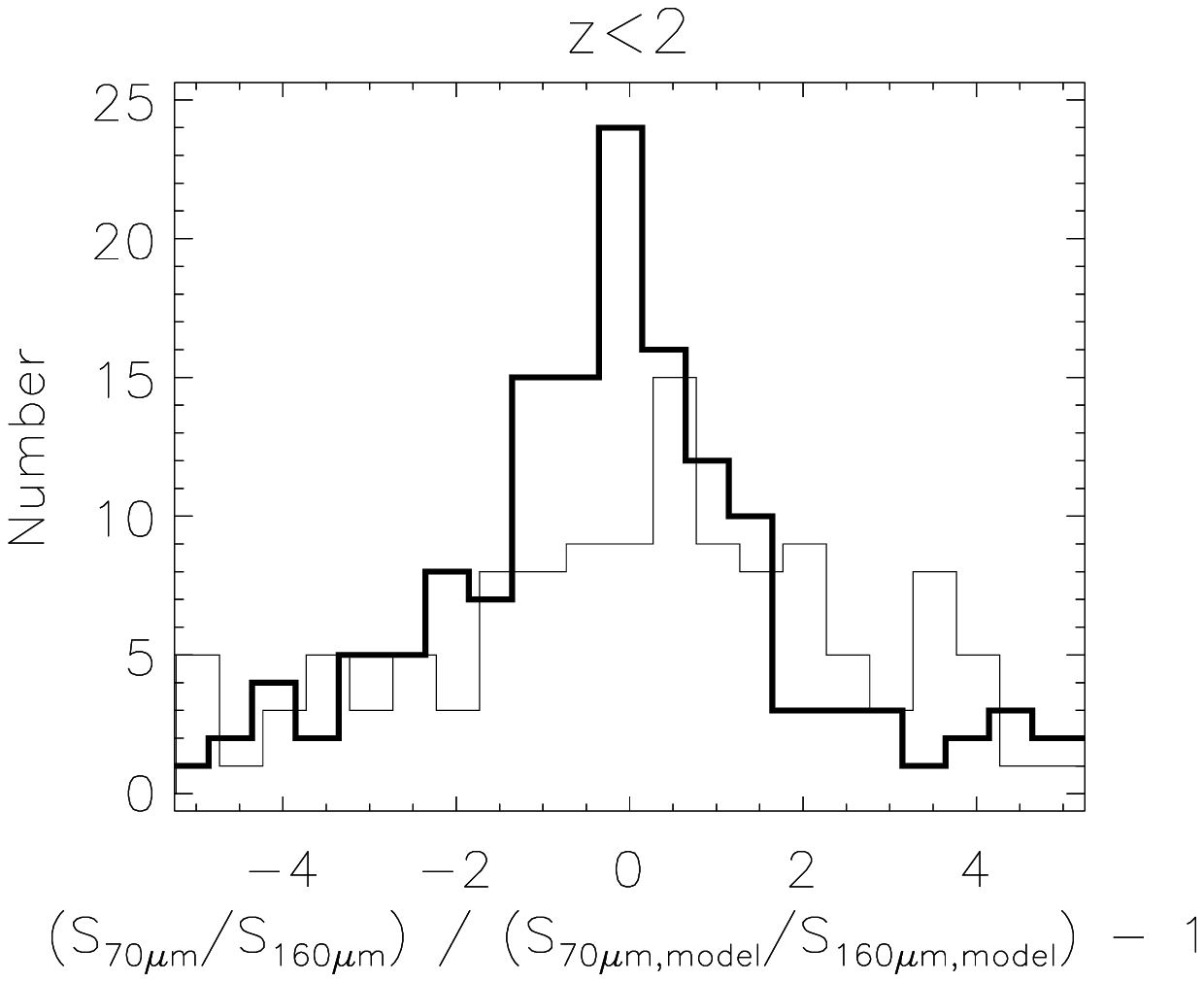}
  \vspace*{-7.2cm}
  \ForceWidth{10.0cm}
  \hSlide{4cm}
  \BoxedEPSF{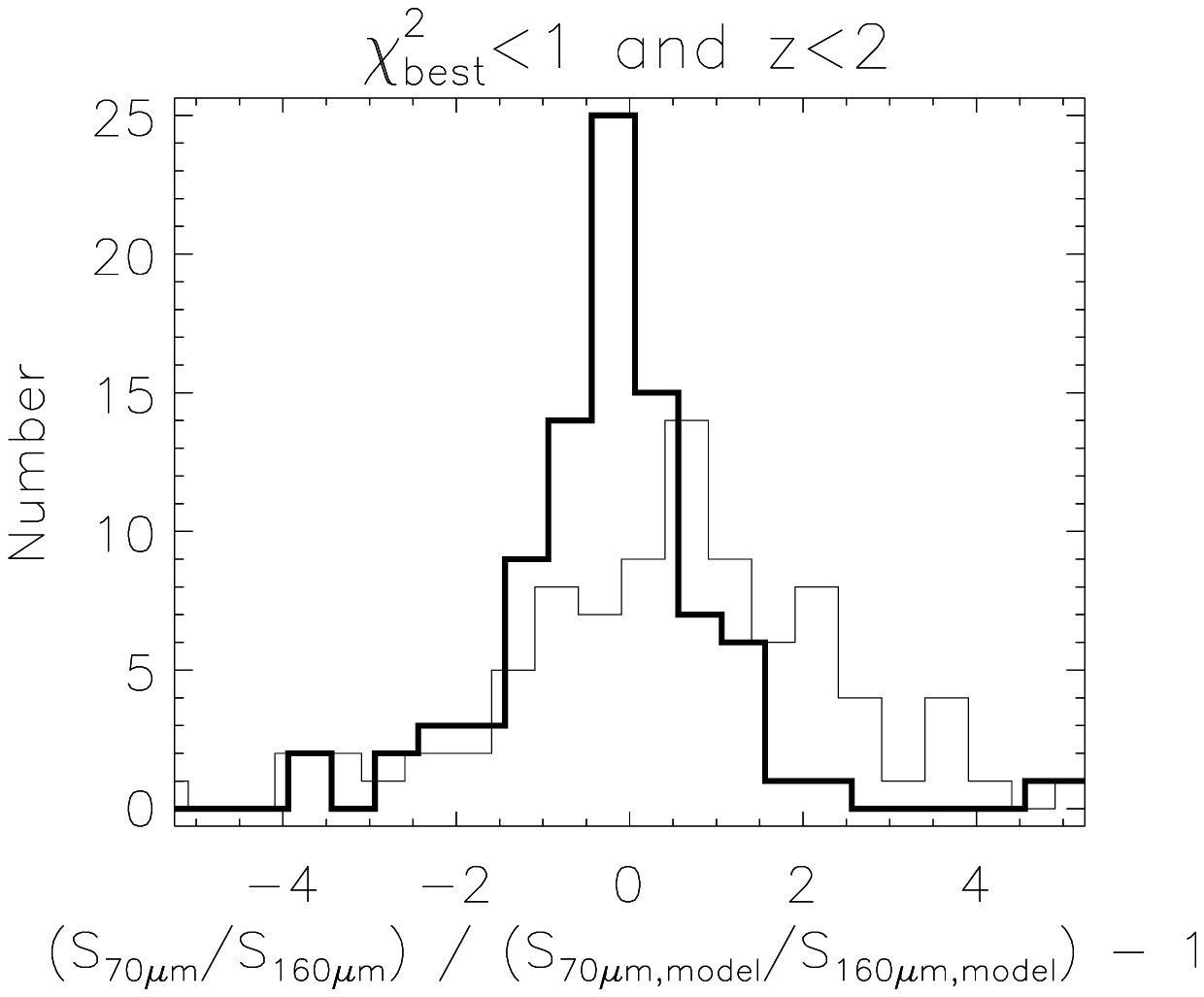}
\caption{\label{fig:chi2}Comparison of the 70$\,\mu$m:160$\,\mu$m flux
  ratio for SWIRE SDSS quasars (including individual SWIRE
  non-detections) for an M82 SED (thick histogram) and an Arp 220 SED
  (thin histogram). The left panel is restricted to objects at
  redshifts $z<2$, and the right panel has the additional restriction
  that the best $\chi^2$ of the M82 and Arp 220 models is better than 1. The
  $x$-axis shows the observed flux ratio divided by the predicted ratio,
  minus one. If the SED model choice is correct on average, the
  histogram should be centred on zero. The M82 histogram (thick)
  appears better centred on zero, so is favoured by this test.}
\end{figure*}

We also tried estimating comparing the 70$\,\mu$m:160$\,\mu$m flux
ratio with the predictions for redshifted Arp 220 and M82 SEDs. The
results are shown in figure \ref{fig:chi2}. If an SED model is correct
on average, the histogram in figure \ref{fig:chi2} should be centred
around zero. This again suggests that an M82 SED is a better match to
the {\it average} quasar far-infrared-submm SED than Arp 220 (though the
most luminous may nevertheless more resemble Arp 220). This is in
contrast with the bright quasars studied by Hatziminaoglou et
al. (2008), though in keeping with their suggestion that fainter
quasars have less heavily obscured SEDs. Note however that we have
already excluded SED dependence on optical luminosity, redshift or
black hole mass as explanations for the trends in figures
\ref{fig:stacked_luminosity_vs_i}, \ref{fig:stacked_luminosity_vs_z},
\ref{fig:stacked_luminosity_vs_bh} and
\ref{fig:stacked_luminosity_vs_z_bh}.

\section{Discussions}
\subsection{Predictions for Herschel}
Stacking analyses in general only yield information on the mean
fluxes, and yield little information on the dispersion within the
population. However, in the case considered here, we find evidence for
a subset of quasars with bright far-infrared fluxes up to ten times
the mean fluxes within the population.

The 40 beams/source confusion limits predicted for Herschel by
Rowan-Robinson (2001) are $4.6\,$mJy at 70$\,\mu$m and $59\,$mJy at
160$\,\mu$m. If we accept this estimate, then most of the targets in
the AGN survey would therefore be very challenging for PACS direct
detection. However, HSPOT reports a $5\sigma$ confusion limit of only
$1.2\,$mJy at $110\,\mu$m, so this forecast may be pessimistic.

We used a crude fit to the data in table \ref{tab:m82_stacks} to
predict the individual SDSS quasar fluxes (figures
\ref{fig:stacked_luminosity_vs_i} and
\ref{fig:stacked_luminosity_vs_z}), from which we estimate that the
Herschel ATLAS survey will detect 92\% of SDSS quasars at $z<0.2$
(though the $z<0.2$ quasars represent only 0.5\% of all SDSS
quasars). At higher redshifts, only the far-infrared-loud subset will
be detectable. We estimate that 66\% of SDSS QSOs with far-infrared
luminosities $5\times$ larger than the mean will be detectable in this
survey, corresponding to about $221(f_5/0.05)$ quasars detected over
$\simeq500$\,deg$^2$, where $f_5$ is the fraction of quasars with
luminosities $5\times$ larger than the mean. The detected fraction of
$5\times$ over-luminous quasars at $z>3$ is only 5\%. However, the
Herschel ATLAS survey should detect about 98\% of all the SDSS quasars
with luminosities $10\times$ larger than the mean, i.e. about
$333(f_{10}/0.05)$ quasars over $500\,$deg$^2$ where $f_{10}$ the
fraction with luminosities $10\times$ larger than the mean.

We have neglected type 2 AGN in this analysis. These will double or
triple the total number of AGN detected by the Herschel ATLAS
survey. AGN not detected individually in this survey will be
detectable in stacking analyses. It will be illuminating to test
whether sub-classes of quasars have a greater tendency to be
far-infrared-loud in this survey (e.g. broad absorption line quasars,
nitrogen-rich quasars).

\subsection{Physical interpretation}
It might be possible for AGN dust tori models to account for the
far-infrared and submm luminosities of quasars, but only by assuming
very high equatorial optical depths and large physical sizes. If
quasar heating dominated the far-infrared outputs throughout our
sample, we would expect a linear correlation between quasar absolute
magnitude and far-infrared luminosity (figure
\ref{fig:stacked_luminosity_vs_i}), whereas the observed correlation
is shallower. While we cannot exclude AGN heating in a subset of our
objects, we will follow Efstathiou \& Rowan-Robinson (1995) in
treating the far-infrared and submm luminosities of quasars as being
typically dominated by star formation.

The far-infrared luminosities scale linearly with the star formation rates as 
\begin{equation}
SFR=\frac{L_{\rm FIR}}{5.8\times10^9L_\odot}M_\odot/{\rm yr}
\end{equation} 
where $L_{\rm FIR}$ is the bolometric luminosity from the starburst
(see Kennicutt 1998), assuming a Salpeter initial mass function (IMF)
from 0.1 to $100\,M_\odot$, implying that our quasars are forming
stars at around $200-2000\,M_\odot$/yr. The local spheroid associated
with a $10^8$ ($10^9$) $M_\odot$ black hole has a mass of about
$5\times10^{10}$ ($5\times10^{11}$) $M_\odot$ (Marconi \& Hunt 2003,
H\"{a}ring \& Rix 2004), though there are indications that the
spheroids are around a factor of 4 less massive at $z=2$ (e.g. McLure
et al. 2006). Sustained star formation at these rates could assemble
the $z=0$ stellar mass hosts in a few $\times10^7$ to a few
$\times10^9$ years.

The correlation between inferred black hole accretion and star
formation is similar to the one reported by Hao et al. (2007), though
they combined low luminosity - low redshift quasars with high
luminosity - high redshift quasars, so their results could also have
been attributable to evolution. We span a much bigger range of the
optical luminosity - redshift parameter space (figure \ref{fig:iz}), so these
caveats do not apply to our results.

The quasar magnitudes in our lowest redshift bin may in principle be
contaminated by the host galaxies. If so, correcting for this effect
would only strengthen the dependence of star formation on quasar
luminosity, since the correction would apply preferentially at the
faintest optical magnitudes.

One difficulty in the interpretation of these results is the
possibility of luminosity-dependent reddening of quasars. However,
even an $A_V$ of one at the faintest end and zero at the brightest
would have little impact on our correlations, given the size of our
errors and range of absolute magnitudes. We have assumed a single
optical spectral index for our quasars, but a slightly better approach
would be to use the optical spectra themselves, correcting for dust
absorption using the Balmer decrement where possible, or using
rest-frame hard X-ray luminosities. Alexander et al. (2005) found an
approximately linear relationship between hard X-ray and far-infrared
luminosities in a heterogeneous sample of AGN-dominated submm-selected
galaxies, though if one adds the submm galaxies classified as
starbursts, their correlation is shallower with a wide dispersion. We
have not excluded candidate starburst-dominated objects. Alexander et
al. (2005) also demonstrate that local AGN show a large scatter in
their star formation - black hole accretion relationship; our
lowest-redshift quasars in figure \ref{fig:stacked_luminosity_vs_i}
show marginal evidence for a steeper correlation than at higher redshifts,
broadly in agreement with these observations of local active galaxies.

The implicit correlations in figure \ref{fig:stacked_luminosity_vs_i}
between star formation rate and black hole accretion rate hint at
common physical parameters (such as gas supply), despite the disparity
of spatial scales, but in keeping with qualitative expectations from
the black hole - spheroid connections in local galaxies
(e.g. Maggorian et al. 1998, Ferrarese \& Merritt 2000). There is no
insight to be gained by supposing the only common parameter is the
total mass of the system, i.e. that this is simply reflecting only a size
dependence, because one must still hypothesize some mechanisms to tie
both parameters to the total mass (e.g. Serjeant et al. 1998); in any
case, the trends in figure \ref{fig:stacked_luminosity_vs_i} follow
approximately $L_{\rm FIR}\propto L_{\rm opt}^{0.5}$ rather than a
linear relationship. Furthermore, the lack of any obvious correlation
with black hole mass at redshifts $0.5<z<4$ (see below) argues against
any simple scaling with the size of the system, at least outside the
local Universe. 

This non-linear relationship and its evolution do not follow
expectations from some semi-analytic models. According to the model of
Croton (2006), the ratio of black hole accretion rate and star
formation rate is constant with scale but increases with
redshift. This is partly due to increased disk disruption at high
redshifts generating starbursts but not black hole accretion, and
partly due to evolution in the black hole feeding rate in this
model. The model does succeed in reproducing the evidence for
evolution in the black hole - bulge mass relationships and
(qualitatively at least) our evolving normalisation of the black hole
accretion -- star formation relation from $z<0.5$ to $1<z<2$. However,
it does not reproduce our scale dependence. Our observations may prove
to be a useful constraint on AGN feedback models.

If AGN feedback directly regulates stellar mass assembly in the host,
then we may expect stronger trends of far-infrared luminosity with
black hole accretion than with black hole mass. Not all of our sample
have black hole mass estimates, so our tests of dependence on black
hole mass and redshift are more noisy than our correlations against
quasar luminosity, though we span a larger logarithmic range of black
hole mass than quasar luminosity. In the local Universe, we see a hint
of a relation between star formation rate and black hole mass (figure
\ref{fig:stacked_luminosity_vs_bh}, table \ref{tab:mbh_fits}). 
At these lower redshifts, the
hosts may already have assembled a large fraction of their $z\simeq0$
stellar masses, so this may partly represent mutual size dependence. 
However, at $0.5<z<4$ there seems to be no evidence for a dependence on
black hole mass (figure \ref{fig:stacked_luminosity_vs_bh}) despite
the dependence on inferred black hole accretion (figure
\ref{fig:stacked_luminosity_vs_i}). 
At $z>4$ there is some evidence for a weak trend with
black hole mass (see table \ref{tab:evol_fits_bh}), varying roughly
as SFR\,$\propto M_{\rm BH}^{1/4}$; the weakness of this trend suggests 
that it is less closely related to the primary underlying physical 
mechanism than the SFR-$L_{\rm opt}$ relation. 
There are nevertheless hints of
trends with redshift at fixed black hole masses (figure
\ref{fig:stacked_luminosity_vs_z}, tables
\ref{tab:m82_bh_stacks} and \ref{tab:evol_fits_bh}). 
More black hole mass estimates are needed to
improve the statistics, but our 
measurements are consistent with
feedback from black hole accretion at $z>1$ regulating the stellar
mass assembly in their hosts.

It is likely that the e-folding timescale for black hole growth
($\tau_{\rm BH}\simeq4\lambda(0.1/\epsilon)\times10^7$\,yr, where
$\lambda$ is the Eddington ratio and $\epsilon$ the accretion
efficiency) is much faster than the stellar mass assembly timescale
(e.g. Malbon et al. 2007). Typical quasar lifetimes at $z<3.5$ may not
be much longer than a single e-folding scale (e.g. Martini \& Weinberg
2001, Shen et al. 2007), though may be several e-foldings at higher
redshifts (Shen et al. 2007). There are 3.6 e-foldings from $I_{\rm
  AB}=-22$ to $I_{\rm AB}=-26$, making it unlikely that the 
relationships in figure \ref{fig:stacked_luminosity_vs_i} 
represent the evolutionary tracks of individual objects. 

We have shown that active nuclei are on average associated with
luminous or ultraluminous starbursts at all redshifts, and all
absolute magnitudes brighter than about $I_{\rm AB}=-22$. This
relationship does not on its own help us address whether the AGN
initiates the starburst or is started concurrently, or whether the AGN
occurs at some midway point, or whether the AGN quenches the
starburst. However, if quasar lifetimes are as long as 600\,Myr at
$z>3.5$, which is the upper limit to the lifetimes suggested by Shen
et al. (2007), then we would expect AGN feedback to have quenched the
star formation in nearly all $z>3.5$ quasars. Our tentative high-z
detections suggest this is not the case. The high-z constraints will
shortly be made much stronger with the large far-infrared and submm
photometric surveys of $z>3.5$ quasars from the Herschel ATLAS key
project, SCUBA-2 and other facilities. Conversely, the shorter
inferred quasar lifetimes at lower redshifts, the lack of evidence for
any dependence of star formation on black hole mass, the observed
dependence of star formation rate with quasar luminosity, and the
local bulge - black hole relationships, are all consistent with
feedback from black hole accretion regulating stellar mass assembly at
lower redshifts.

\section*{Acknowledgments}
This research has made use of the NASA/IPAC Infrared Science Archive
and the NASA/IPAC Extragalactic Database (NED), which are operated by
the Jet Propulsion Laboratory, California Institute of Technology,
under contract with the National Aeronautics and Space
Administration. We would particularly like to thank Anastasia Alexov,
John C. Good and Anastasia Laity at IPAC for their invaluable help
with the IRAS archive. We would also like to thank the anonymous
referee for several helpful suggestions. This research
was supported by STFC grants PP/D002400/1 and PP/E001408.

\appendix

\section[]{Discussion of Palomar-Green quasars}

We use our SCANPI 100$\,\mu$m measurements in preference to other IRAS
100$\,\mu$m measurements (e.g. Saunders et al. 1989), but that
withstanding, we use the 100$\,\mu$m photometry with the smallest
errors with the exceptions of the cases discussed below. At 60$\,\mu$m
we use the measurements with the smallest errors regardless of their
source, again with the exceptions discussed below. Haas et al. (2000,
2003) do not quote errors on their ISO photometry, but state that the
detections range from $3-10\sigma$. We conservatively assume $3\sigma$
unless stated otherwise. Objects dominated by non-thermal emission at
60-100$\,\mu$m have been eliminated from our stacking analyses. The
adopted photometry for the Palomar-Green sample is given in 
table \ref{tab:pg_photometry}. 

0007+106, or MRK1501, is a radio-loud flat-spectrum quasar, and is
likely to have non-thermal emission dominating at 60-100$\,\mu$m.

0050+124, or 1Zw1, has a 60 (100) $\mu$m measurement from ISO of 1752
(2339) mJy reported by Haas et al. 2000, but these are inconsistent
with our SCANPI measurements of 2161$\pm$52 (1749$\pm$187) mJy which show
no obvious anomalies in the fits. We have opted to use our SCANPI
photometry.

0157+001 has a 60$\,\mu$m flux measurement of 2210\,mJy reported by
Haas et al. 2003, though errors are not quoted. Sanders et al. 1989
quotes an IRAS measurement of 2377$\pm$56\,mJy. Our IRAS SCANPI
measurement of 2348$\pm$73\,mJy is consistent with the latter rather than
the former. We have opted to use our SCANPI photometry, which shows no
obvious anomalies in the fit.

0832+251 (z=0.320) is reported as $<126$\,mJy at 60$\,\mu$m in Sanders et
al. 1989 but has a SCANPI measurement of 352$\pm$60\,mJy.  There is a
similar discrepancy at 100$\,\mu$m. However this is because SCANPI's
fit is strongly affected by the nearby IRAS galaxy IRAS 08325+2512 at
z=0.017, which is 2.5 arcmin from the QSO. Both the 60 and 100$\,\mu$m
coadded scans appear to be fairly flat off-source. We therefore found
the maximum-likelihood fit for the amplitudes of a source fixed at the
target position, and another with a position allowed to vary.

0838+770 has an ISO 100$\,\mu$m flux measurement of 180\,mJy from Haas
et al. 2003, which disagrees with the Sanders et al. 1989 IRAS
measurement of 426$\pm$30\,mJy. Our IRAS SCANPI measurement is
293$\pm$184\,mJy, though with strong baseline drifts in the coadded
timeline. Given the uncertainties we have uncovered in the IRAS
100$\,\mu$m flux calibration at this level, and the baseline drifts in
our SCANPI data, we have opted to use the Haas et al. photometry with
an assumed 33\% error.

1001+054 has an ISO 60$\,\mu$m flux measurement of 140\,mJy from Haas et
al. 2003, which disagrees with our IRAS SCANPI measurement of
23$\pm$49\,mJy. There are no hints of flux at the target position in our
coadded scans. Sanders et al. 1989 report 27$\pm$9\,mJy, which has a
remarkably small quoted error. Nothing is reported at this position in
either the IRAS PSC or FSC, and nothing is evident on the ISSA
plates. We have opted to use the Sanders photometry.

1022+519 has a 100$\,\mu$m flux measurement from the IRAS Faint Source
Reject catalogue of 798$\pm$176\,mJy. Our SCANPI photometry is
200$\pm$103\,mJy, with a fairly stable coadded baseline. Owing to
this, and the lower quoted error of our SCANPI measurements, together
with the uncertainties we have uncovered in the catalogued IRAS 100
micron fluxes at this level, we have opted to use our SCANPI
measurement.

1100+772, or 3C249.1, is radio-loud and probably synchrotron-dominated
at both 60$\,\mu$m and 100$\,\mu$m.

1103-006 has a 60$\,\mu$m IRAS flux of 130$\pm$51\,mJy quoted in Sanders
et al. 1989, while our SCANPI measurement is -8$\pm$98\,mJy, though the
coadded scans are affected by baseline drifts. We have used the
Sanders measurement.

1211+143 The Haas et al. 60$\,\mu$m measurement of 518\,mJy disagrees
both with the Sanders et al. 1989 measurement of 305$\pm$53\,mJy and
our own IRAS SCANPI measurement of 284$\pm$81\,mJy. At 100$\,\mu$m the
disagreement is more striking, with Haas et al. reporting an upper
limit of $<279$\,mJy, while our SCANPI measurement is 427$\pm$182\,mJy
and Sanders et al. report 689$\pm$119\,mJy. Although baseline drifts
are evident in our coadded scans, the background subtractions at the
position of our target appear to be reliable, and the profile is
well-fit. We have opted to use the lowest noise IRAS measurements.

1226+023, or 3C273, has a clearly non-thermal spectrum in IRAS
passbands.

1302-102 is radio-loud and probably synchrotron-dominated at 60 and
100$\,\mu$m.

1351+640 has a 100$\,\mu$m flux measurement from Haas et al. 2003 of
526\,mJy, but the IRAS SCANPI measurement is 912$\pm$156\,mJy. Although
baseline drifts are apparent in our coadded scans, the background
subtractions at the position of our target appear to be
reliable. Sanders et al. 1989 report 1184$\pm$26\,mJy. We have opted to
use our SCANPI measurement.

1501+106 has a 60$\,\mu$m measurement from Haas et al. 2003 of 750\,mJy,
but our SCANPI measurement is 473$\pm$36\,mJy. Sanders et al. 1989 report
486$\pm$42\,mJy. We adopt our SCANPI measurement as the most likely
lowest-noise choice.

1545+210 is radio-loud and probably synchrotron-dominated at 60 and
100$\,\mu$m.

1613+658 has a 100$\,\mu$m measurement from Haas et al. 2000 of
1002\,mJy, consistent with the Sanders et al. 1989 measurement of
1090$\pm$59\,mJy, but our SCANPI measurement is 474$\pm$70\,mJy. Our
SCANPI measurement shows a baseline drift that may be over-corrected,
so we opt to use the Haas et al. measurement, and assume an error of
10\%.

1704+608, or 3C351, is radio-loud and probably synchrotron-dominated
at 60 and 100$\,\mu$m.

1718+481 is radio-loud and probably synchrotron-dominated at 60 and
100$\,\mu$m.

2209+184 is radio-loud and probably synchrotron-dominated at 60 and
100$\,\mu$m.

2251+113 is radio-loud and probably synchrotron-dominated at 60 and
100$\,\mu$m.

2344+092 is radio-loud and probably synchrotron-dominated at 60 and
100$\,\mu$m.

2349-014 is radio-loud and probably synchrotron-dominated at 60 and
100$\,\mu$m.

Only contains objects which are not dominated by synchrotron at 60-100um 

\begin{table*}
\begin{tabular}{llllllll}
  Name	 &      Right Ascension	& Declination                   & $S_{60}$      & $S_{100}$ & Redshift\\
  	 &      (J2000)	        & (J2000)                       & (mJy)         & (mJy) &  \\
0002+051 &	00 05 20.2155 	& +05 24 10.800 	        & $15\pm$58	& $-43\pm$232	& 1.900\\
0003+158 &	00 05 59.200  	& +16 09 48.00 	        	& $37 \pm$70	& $-75\pm$378	& 0.45\\
0003+199 &	00 06 19.521    & +20 12 10.49 	        	& $260\pm$93	& $-509\pm$781	& 0.025\\
0026+129 &	00 29 13.6 	& +13 16 03 	        	& $0.9 \pm$109	& $-437\pm$105	& 0.142\\
0043+039 &	00 45 47.3 	& +04 10 24 	        	& $-2.7 \pm$48	& $165\pm$154	& 0.385\\
0044+030 &	00 47 05.91  	& +03 19 55.0 	        	& $70 \pm$49	& $158\pm$79	& 0.623\\
0049+171 &	00 51 54.800  	& +17 25 58.40 	        	& $16 \pm$73	& $642\pm$286	& 0.064\\
0050+124 &	00 53 34.940  	& +12 41 36.20 	        	& $2161\pm$52	& $1749\pm$187	& 0.061\\
0052+251 &	00 54 52.1 	& +25 25 38 	        	& $93 \pm$18	& $163\pm$54	& 0.155\\
0117+213 &	01 20 17.2 	& +21 33 46 	        	& $-0.7 \pm$71	& $-96\pm$173	& 1.493\\
0119+229 &	01 22 40.58 	& +23 10 15.1 	        	& $921 \pm$63	& $773\pm$264	& 0.053\\
0157+001 &	01 59 50.211  	& +00 23 40.62  	        & $2348 \pm$73	& $1915\pm$168	& 0.163\\
0804+761 &	08 10 58.600  	& +76 02 42.00 	        	& $191 \pm$42	& $121\pm$36	& 0.1\\
0832+251 &	08 35 35.820    & +24 59 40.65 	        	& $182\pm$17	& $194\pm$68	& 0.320\\
0838+770 &	08 44 45.26  	& +76 53 09.5 	        	& $174 \pm$9	& $180\pm$60	& 0.131\\
0844+349 &	08 47 42.4 	& +34 45 04 	        	& $163 \pm$41	& $178\pm$337	& 0.064\\
0906+484 &	09 10 10.010    & +48 13 41.80 	        	& $172\pm$10	& $210\pm$134	& 0.118\\
0921+525 &	09 25 12.870    & +52 17 10.52 	        	& $131\pm$52	& $-171\pm$120	& 0.035\\
0923+201 &	09 25 54.700  	& +19 54 05.00 	        	& $271 \pm$58	& $858\pm$291	& 0.190\\
0923+129 &	09 26 03.292 	& +12 44 03.63 	        	& $590.1\pm$53	& $675\pm$239	& 0.029\\
0931+437 &	09 35 02.540  	& +43 31 10.70 	        	& $107\pm$72	& $110\pm$113	& 0.457\\
0934+013 &	09 37 01.030 	& +01 05 43.48 	        	& $190\pm$102	& $-30\pm$673	& 0.050\\
0935+417 &	09 38 57.00  	& +41 28 20.79 	        	& $28\pm$56	& $-3.1\pm$156	& 1.980 \\
0946+301 &	09 49 41.113  	& +29 55 19.24 	        	& $36 \pm$45	& $-68\pm$116	& 1.216\\
0947+396 &	09 50 48.380  	& +39 26 50.50 	        	& $201 \pm$47	& $279\pm$137	& 0.206\\
0953+414 &	09 56 52.4 	& +41 15 22 	        	& $107 \pm$56	& $47\pm$117	& 0.239\\
1001+054 &	10 04 20.140  	& +05 13 00.50 	        	& $27 \pm$9	& $146\pm$49	& 0.161\\
1004+130 &	10 07 26.100  	& +12 48 56.20 	        	& $191 \pm$42	& $0.5\pm$130	& 0.24\\
1008+133 &	10 11 10.857  	& +13 04 11.90 	        	& $58 \pm$58	& $-91\pm$224	& 1.287\\
1011-040 &	10 14 20.69  	& -04 18 40.5 	        	& $163 \pm$42	& $-53\pm$153	& 0.058\\
1012+008 &	10 14 54.900  	& +00 33 37.30 	        	& $-25 \pm$74	& $-23\pm$348	& 0.185\\
1022+519 &	10 25 31.278 	& +51 40 34.87 	        	& $153 \pm$40	& $200\pm$103	& 0.045\\
1048+342 &	10 51 43.900  	& +33 59 26.70	        	& $7.3 \pm$70	& $-9.7\pm$168	& 0.167\\
1048-090 &	10 51 29.900  	& -09 18 10.00 	        	& $69 \pm$60	& $215\pm$389	& 0.344\\
1049-005 &	10 51 51.450  	& -00 51 17.70  	        & $191 \pm$56	& $-11\pm$255	& 0.357\\
1103-006 &	11 06 31.775  	& -00 52 52.47 	        	& $130 \pm$51	& $-234\pm$241	& 0.425\\
1112+431 &	11 15 06.020 	& +42 49 48.90 	        	& $182\pm$47	& $132\pm$117	& 0.302\\
1114+445 &	11 17 06.400  	& +44 13 33.30 	        	& $191 \pm$47	& $200\pm$60	& 0.144\\
1115+080 &	11 18 16.950  	& +07 45 58.20 	        	& $769 \pm$96	& $871\pm$242	& 1.722\\
1115+407 &	11 18 30.290  	& +40 25 54.00 	        	& $265 \pm$71	& $143\pm$162	& 0.154\\
1119+120 &	11 21 47.103  	& +11 44 18.26 	        	& $452 \pm$49	& $481\pm$267	& 0.049\\
1121+422 &	11 24 39.190  	& +42 01 45.00 	        	& $-54 \pm$77	& $66\pm$237	& 0.234\\
1126-041 &	11 29 16.661  	& -04 24 07.59 	        	& $669 \pm$26	& $415\pm$686	& 0.06\\
1138+040 &	11 41 16.530  	& +03 46 59.60 	        	& $-1.1\pm$54	& $-69\pm$151	& 1.876\\
1148+549 &	11 51 20.460  	& +54 37 33.10 	        	& $213 \pm$32	& $239\pm$78	& 0.969\\
1149-110 &	11 52 03.544  	& -11 22 24.32 	        	& $215 \pm$67	& $314\pm$105	& 0.049\\
1151+117 &	11 53 49.270  	& +11 28 30.40 	        	& $137 \pm$70	& $218\pm$147	& 0.176\\
\end{tabular}
\caption{\label{tab:pg_photometry}This table lists the adopted photometry for the Palomar-Green quasar
sample. The columns give the name, the position, the $60\,\mu$m and $100\,\mu$m photometry,
and the redshift. (Continued overleaf.)}
\end{table*}
\begin{table*}
\begin{tabular}{llllllll}
  Name	 &      Right Ascension	& Declination                   & $S_{60}$      & $S_{100}$ & Redshift\\
  	 &      (J2000)	        & (J2000)                       & (mJy)         & (mJy) & \\
1202+281 &	12 04 42.1 	& +27 54 11 	        	& $176 \pm$41	& $154\pm$133	& 0.165\\
1206+459 &	12 08 58.012  	& +45 40 35.87 	        	& $215 \pm$64	& $383\pm$89	& 1.158\\
1211+143 &	12 14 17.7 	& +14 03 13 	        	& $305 \pm$53	& $427\pm$182	& 0.084\\
1216+069 &	12 19 20.9 	& +06 38 38 	        	& $48 \pm$60	& $150\pm$144	& 0.331\\
1222+228 &	12 25 27.4 	& +22 35 13 	        	& $67 \pm$65	& $-78\pm$159	& 2.046\\
1229+204 &	12 32 03.605  	& +20 09 29.21 	        	& $154 \pm$64	& $317\pm$105	& 0.063\\
1241+176 &	12 44 10.859  	& +17 21 04.32 	        	& $132 \pm$55	& $217\pm$72	& 1.273\\
1244+026 &	12 46 35.240  	& +02 22 08.70 	        	& $280 \pm$51	& $362\pm$121	& 0.048\\
1247+267 &	12 50 05.7 	& +26 31 08 	        	& $102 \pm$55	& $174\pm$58	& 2.038\\
1248+401 &	12 50 48.368  	& +39 51 39.80 	        	& $224 \pm$51	& $-53\pm$190	& 1.03\\
1254+047 &	12 56 59.959  	& +04 27 34.16 	        	& $98 \pm$51	& $242\pm$81	& 1.024\\
1259+593 &	13 01 12.930  	& +59 02 06.70 	        	& $34 \pm$51	& $-6.5\pm$125	& 0.478\\
1307+085 &	13 09 47.0 	& +08 19 49 	        	& $117 \pm$51	& $155\pm$52	& 0.155\\
1309+355 &	13 12 17.767  	& +35 15 21.24 	        	& $147 \pm$46	& $-29\pm$104	& 0.184\\
1310-108 &	13 13 05.8 	& -11 07 42 	        	& $102 \pm$77	& $29\pm$288	& 0.035\\
1322+659 &	13 23 49.5 	& +65 41 48 	        	& $90 \pm$30	& $100\pm$33	& 0.168\\
1329+412 &	13 31 41.130  	& +41 01 58.70 	        	& $136 \pm$58	& $123\pm$131	& 1.93\\
1333+176 &	13 36 02.0 	& +17 25 13 	        	& $121 \pm$53	& $157\pm$168	& 0.554\\
1338+416 &	13 41 00.780  	& +41 23 14.10 	        	& $30 \pm$45	& $125\pm$149	& 1.219\\
1341+258 &	13 43 56.7 	& +25 38 48 	        	& $84\pm$40	& $527\pm$185	& 0.087\\
1351+236 &	13 54 06.432  	& +23 25 49.09 	        	& $364 \pm$51	& $306\pm$192	& 0.055\\
1351+640 &	13 53 15.808  	& +63 45 45.41 	        	& $757 \pm$8	& $912\pm$156	& 0.088\\
1352+183 &	13 54 35.6 	& +18 05 17 	        	& $-85 \pm$47	& $-29\pm$218	& 0.152\\
1352+011 &	13 54 58.7 	& +00 52 10 	        	& $104 \pm$59	& $109\pm$165	& 1.121\\
1354+213 &	13 56 32.7 	& +21 03 52 	        	& $-15 \pm$54	& $-84\pm$158	& 0.3\\
1402+261 &	14 05 16.195  	& +25 55 34.93 	        	& $318 \pm$47	& $213\pm$71	& 0.164\\
1404+226 &	14 06 21.8 	& +22 23 46 	        	& $51 \pm$52	& $-42\pm$164	& 0.098\\
1407+265 &	14 09 23.9 	& +26 18 21 	        	& $171 \pm$51	& $-16\pm$126	& 0.94\\
1411+442 &	14 13 48.3 	& +44 00 14 	        	& $162 \pm$17	& $140\pm$47	& 0.09\\
1415+451 &	14 17 00.820  	& +44 56 06.40 	        	& $112 \pm$37	& $147\pm$49	& 0.114\\
1416-129 &	14 19 03.800  	& -13 10 44.00 	        	& $30 \pm$67	& $198\pm$398	& 0.129\\
1425+267 &	14 27 35.540  	& +26 32 13.61 	        	& $79 \pm$58	& $-16\pm$144	& 0.366\\
1426+015 &	14 29 06.588  	& +01 17 06.48 	        	& $318\pm$47	& $62\pm$102	& 0.086\\
1427+480 &	14 29 43.070  	& +47 47 26.20 	        	& $82 \pm$25	& $92\pm$31	& 0.221\\
1435-067 &	14 38 16.1 	& -06 58 21 	        	& $-16 \pm$75	& $-229\pm$233	& 0.126\\
1440+356 &	14 42 07.463  	& +35 26 22.92 	        	& $652 \pm$21	& $793\pm$87	& 0.079\\
1444+407 &	14 46 45.940  	& +40 35 05.70 	        	& $57 \pm$30	& $80\pm$27	& 0.267\\
1448+273 &	14 51 08.8 	& +27 09 27 	        	& $117 \pm$37	& $-34\pm$100	& 0.065\\
1501+106 &	15 04 01.201  	& +10 26 16.15 	        	& $473 \pm$36	& $77\pm$144	& 0.036\\
1512+370 &	15 14 43.042  	& +36 50 50.41 	        	& $61 \pm$20	& $160\pm$159	& 0.37\\
1519+226 &	15 21 14.2 	& +22 27 43 	        	& $-21 \pm$49	& $155\pm$150	& 0.137\\
1522+101 &	15 24 24.6 	& +09 58 30 	        	& $37 \pm$71	& $-38\pm$201	& 1.321\\
1534+580 &	15 35 52.361  	& +57 54 09.21 	        	& $140 \pm$51	& $136\pm$128	& 0.03\\
1535+547 &	15 36 38.361  	& +54 33 33.21 	        	& $61 \pm$32	& $81\pm$148	& 0.038\\
1538+477 &	15 39 34.8  	& +47 35 31 	        	& $97\pm$39	& $107\pm$121	& 0.770\\
1543+489 &	15 45 30.240  	& +48 46 09.10 	        	& $348 \pm$26	& $371\pm$79	& 0.4\\
1552+085 &	15 54 44.6 	& +08 22 22 	        	& $-82 \pm$103	& $-276\pm$140	& 0.119\\
1612+261 &	16 14 13.210  	& +26 04 16.20 	        	& $252 \pm$72	& $330\pm$629	& 0.131\\
1613+658 &	16 13 57.179  	& +65 43 09.58 	        	& $635 \pm$19	& $1002\pm$100	& 0.129\\
1617+175 &	16 20 11.288  	& +17 24 27.70  	        & $52 \pm$45	& $45\pm$108	& 0.114\\
1626+554 &	16 27 56.0 	& +55 22 31 	        	& $-28 \pm$46	& $70\pm$23	& 0.133\\
1630+377 &	16 32 01.120  	& +37 37 50.00 	        	& $5.9 \pm$36	& $-105\pm$110	& 1.466\\
1634+706 &	16 34 28.884  	& +70 31 33.04 	        	& $318 \pm$23	& $444\pm$80	& 1.334\\
1700+518 &	17 01 24.800 	& +51 49 20.00 	        	& $480 \pm$36	& $374\pm$125	& 0.292\\
1715+535 &	17 16 35.5  	& +53 28 15 	        	& $3.2\pm$60	& $-88\pm$123	& 1.920\\
2112+059 &	21 14 52.6 	& +06 07 42 	        	& $105 \pm$19	& $193\pm$370	& 0.466\\
2130+099 &	21 32 27.813  	& +10 08 19.46 	        	& $479 \pm$12	& $485\pm$162	& 0.062\\
2214+139 &	22 17 12.26  	& +14 14 20.9 	        	& $337 \pm$11	&  	       	& 0.067\\
2233+134 &	22 36 07.680  	& +13 43 55.30 	        	& $80 \pm$68	& $-647\pm$302	& 0.325\\
2302+029 &	23 04 45.0 	& +03 11 46 	        	& $130 \pm$66	& $118\pm$174	& 1.044\\
2304+042 &	23 07 02.9  	& +04 32 57 	        	& $60\pm$63	& $70\pm$130	& 0.042\\
2308+098 &	23 11 17.758  	& +10 08 15.46 	        	& $83\pm$87	& $-539\pm$420	& 0.433\\
\end{tabular}
\caption{\label{tab:pg_photometry2}(continuation of table \ref{tab:pg_photometry}.) 
This table lists the adopted photometry for the Palomar-Green quasar
sample. The columns give the name, the position, the $60\,\mu$m and $100\,\mu$m photometry,
and the redshift.}
\end{table*}

\end{document}